\documentclass[final]{nst}

\usepackage{subfigure,dcolumn}
\usepackage[T2A,T1]{fontenc}
\usepackage[russian,english]{babel}

\usepackage{listings}
\begin{document}

\title{The reduced-width amplitude in nuclear cluster physics}\thanks{Supported by the National Key R$\&$D Program of China (2023YFA1606701) and the National Natural Science Foundation of China under contract Nos. 12175042 and 12147101.
}

\author{De-Ye Tao}
\affiliation{Key Laboratory of Nuclear Physics and Ion-beam Application (MOE), Institute of Modern Physics, Fudan University, Shanghai 200433, China}

\author{Bo Zhou}
\email[Corresponding author, ]{zhou_bo@fudan.edu.cn}
\affiliation{Key Laboratory of Nuclear Physics and Ion-beam Application (MOE), Institute of Modern Physics, Fudan University, Shanghai 200433, China}
\affiliation{Shanghai Research Center for Theoretical Nuclear Physics, NSFC and Fudan University, Shanghai 200438, China}

\begin{abstract}
The reduced-width amplitude, as a cluster overlap amplitude, is one important physical quantity for analyzing clustering in the nucleus depending on specified channels and has been calculated and applied widely in nuclear cluster physics.
In this review, we briefly revisit the theoretical framework for calculating the reduced-width amplitude, as well as the outlines of cluster models to obtain the microscopic or semi-microscopic cluster wave functions.
Besides, we also introduce recent progress related to the cluster overlap amplitudes, such as implementation for cross-section estimation and extension to three-body clustering analysis. 
Comprehensive examples are provided to show the applications of the reduced-width amplitude in analyzing cluster structures.
\end{abstract}

\keywords{nuclear clustering, microscopic cluster model, reduced-width amplitude}

\maketitle

\section{Introduction}

Clustering is one of the most important aspects in understanding the nuclear structure~\cite{freer2018RMP035004,zhou2019FoP14401}.
In both stable and unstable exotic nuclei, clustering fundamentally affects the structure and reaction features. 
Since the 1960s, a variety of cluster models~\cite{funaki2015PPNP78,Descouvemont2012} have been proposed and applied to study the exotic nuclear structures that cannot be understood from the pure shell-model perspective. 
Nowadays, clustering effects play increasingly important roles not only in nuclear structure but also in many other fields of nuclear physics, such as heavy-ion collisions~\cite{cao2023PRC064906,xu2018NST186,shi2021NST66}, astrophysical nucleosynthesis~\cite{sun2024NC1074,pais2018NST181}, and nuclear matter~\cite{tanaka2021S260,ropke2011JPCS012023}.

According to cluster models, nucleons are assigned to different groups to construct clusters, while the relative-motion wave function between the clusters is solved through the equation of motion.
In microscopic cluster models, including the resonating group method~(RGM)~\cite{wheeler1937PR1107,horiuchi1977PoTPS90} and the generator coordinate method~(GCM)~\cite{hill1953PR1102,griffin1957PR311,horiuchi1977PoTPS90}, the antisymmetrization between all the nucleons is guaranteed.
On the other hand, the orthogonality condition model~(OCM)~\cite{saito1968PoTP893,saito1969PoTP705,saito1977PoTPS11}, as a semi-microscopic cluster model, simulates the effect of antisymmetrization by requiring that the inter-cluster wave function must be orthogonal to the forbidden states.
In recent years, a novel kind of wave function, proposed by the four authors Tohsaki, Horiuchi, Schuck, and R\"opke~(THSR)~\cite{tohsaki2001PRL192501} and named after them, was utilized for studying the famous Hoyle state~\cite{hoyle1954AJSS121} and the analogous $4\alpha$ gaslike state of $^{16}\mathrm{O}$.
A decade later, the generalized THSR wave function~\cite{zhou2013PRL262501} was proposed to describe the clustering structure in nuclei~\cite{zhou2014PRC034319,lyu2015PRC014313}.

In nuclear physics, a collection of physical quantities, including the root-mean-square (r.m.s.) radius, the monopole transition strength $M(E0)$, and the cluster decay width, can reflect the degree of clustering in the nucleus.
A rather straightforward indicator of clustering is the reduced-width amplitude (RWA), or overlap amplitude~\cite{horiuchi1977PoTPS90,kanada-enyo2014PTEP073D02}.
This amplitude is defined as the overlap between the wave function of the nucleus and the cluster-coupling wave function in a specified channel, depending on the distance between the clusters.
As a result, the RWA indicates not only the cluster formation probabilities but also the relative motion between clusters.
The radial nodal excitation of the inter-cluster motion can be inferred from the relation $N=2n+l$, with $N$ denoting the principal quantum number of relative motion, $n$ denoting the radial quantum number equal to the number of nodes exhibited by RWA, and $l$ the quantum number of orbital angular momentum~\cite{wildermuth1977}.
It is worth mentioning that, as a quantity defined in cluster-model analysis, RWA has essentially the same physical meaning as the cluster form factor in the no-core shell model (NCSM)~\cite{navratil2004PRC054324}, and the overlap function in reaction theories~\cite{timofeyuk2014JPGNPP094008}.

Sharing the equivalent definition with the overlap function, the RWA can also serve as an important input parameter in evaluating reaction cross sections, thus providing more precise microscopic structural information for nuclear reaction studies~\cite{descouvemont2022EPJA193,watanabe2021PRCL031601,yoshida2019PRC044601,lyu2018PRC044612,fukui2016PRC034606}. 
In traditional reaction theories such as the distorted-wave Born approximation~(DWBA) method~\cite{satchler1983,glendenning1983DNR45}, the structural information of the participant nuclei is approximated using the simple optical-potential model, and the relative-motion wave function must be normalized by the adjustable spectroscopic factor, which is fitted according to experimental data to include the effects of antisymmetrization and core-excitation~\cite{descouvemont2022EPJA193}. 
Thus, adopting the microscopically obtained RWA as the input in reaction theories would significantly improve the precision and self-consistency of the calculation of cross sections as more detailed information of the participant nuclei and interaction is involved. 

The present article aims to provide a brief yet comprehensible overview of the calculation and application of the RWA.
As important inputs, various cluster-model wave functions, along with outlines of the theoretical models used to obtain them, are introduced in the next section.
In Sec.~\ref{rwa}, we provide the definition and calculation methods of RWA based on the cluster-model wave functions. 
The features and extension to three-body analysis are also discussed.
Applications of RWA in cluster structure analysis are presented in Sec.~\ref{app}.
Finally, in Sec.~\ref{sum} we give the summary and outlooks.

\section{Nuclear cluster wave functions}
\label{cluster_wf}

In light nuclei, the cluster configurations usually evolve with the increase of the excitation energy. 
In general, the ground state exhibits a rather compact structure, while in the excited states, especially close to the break-up thresholds of cluster emission, a variety of interesting cluster structures emerge, as illustrated by the famous Ikeda diagram~\cite{ikeda1968PTPS464}. 

One of the aims of the microscopic cluster models~\cite{freer2018RMP035004} is to describe and understand the correlations between clusters or nucleons in states exhibiting significant clustering effect. 
The microscopic studies on nuclear clustering originated following the proposal of the resonating group method (RGM) by Wheeler in 1937~\cite{wheeler1937PR1107}.
Subsequently, various cluster models were introduced to theoretically analyze the structure and scattering characters of light nuclei.
To describe properly the nuclear system, one should take into account the antisymmetrization effect in the wave function, as a consequence of the indistinguishability of nucleons, and the Hamiltonian of the nuclear system can be expressed as
\begin{equation}
    \mathcal{H}=-\frac{\hbar^2}{2m}\sum_i\nabla_i^2-T_\mathrm{c.m.}+V,
\end{equation}
where $T_\mathrm{c.m.}$ denotes the kinetic energy of the center-of-mass (c.o.m.) motion of the nucleus, and $V$ is the interaction between nucleons, which usually includes the two-body nuclear force, the Coulomb interaction, and the spin-orbit interaction.
Note that the choice of nuclear force is crucial for properly describing the nuclear system.
In microscopic cluster models, effective nucleon-nucleon interactions, such as the Volkov No.~2~\cite{volkov1965NP33} and the Minnesota potential~\cite{thompson1977NPA53}, are frequently adopted, in which the interaction parameters are determined by fitting some fundamental features of the studied system, e.g., the binding energy or the phase shift of $\alpha$ cluster.
On the other hand, as semi-microscopic cluster models disregard the internal structure of clusters, these theories usually treat the two-body nuclear force as a phenomenological cluster-cluster potential or infer it from nucleon-nucleon potentials through a folding procedure.

\subsection{Resonating group method}

The resonating group method~(RGM) was formulated as early as 1937 to microscopically study the scattering between light nuclei. 
In this method, the nucleons are separated into several groups, as the precursor of the concept of ``cluster'', while the exchange effect between identical nucleons from different groups is taken into account as if the nucleon is resonating between each group. 
Since the 1960s, intensive research has been carried out utilizing RGM to analyze the cluster structures of nuclei.
Considering the two-cluster system, $A=C_1+C_2$, the RGM wave function is defined as~\cite{saito1977PoTPS11}
\begin{equation}
    \Psi^\mathrm{RGM}=\mathcal{A}\{\phi_{C_1} \phi_{C_2} \chi(\boldsymbol\rho)\},
\end{equation}
where $\phi_1$ and $\phi_2$ denote the internal wave functions of the clusters $C_1$ and $C_2$, and $\chi(\boldsymbol\rho)$ is the inter-cluster relative wave function, depending on the relative coordinate between the c.o.m.\ of the two clusters $\boldsymbol\rho$.
The antisymmetrization operator $\mathcal{A}$ is applied to the dynamical coordinates of the nucleons between clusters $C_1$ and $C_2$.
The relative wave function $\chi(\boldsymbol\rho)$ can be obtained as the solution of the equation of motion
\begin{equation}
    \int\left[H(\boldsymbol\rho,\boldsymbol\rho')-E N(\boldsymbol\rho,\boldsymbol\rho')\right]\chi(\boldsymbol\rho')d\boldsymbol\rho'=0,
\end{equation}
in which $H(\boldsymbol\rho,\boldsymbol\rho')$ and $N(\boldsymbol\rho,\boldsymbol\rho')$ represent the RGM kernels of Hamiltonian and normalization operators.

\subsection{Generator coordinate method and the Brink wave function}

The actual calculation using the RGM is tedious, requiring solving an integro-differential equation as well as the analytical evaluation of the Hamiltonian and norm kernels. 
Furthermore, the extension to three or more cluster systems within the RGM framework is much more cumbersome.
The proposal of the generator coordinate method (GCM), which is essentially equivalent to RGM~\cite{saito1977PoTPS11,horiuchi1970PTP375}, makes it much easier to perform the calculation and to extend the framework to multi-cluster systems.
The GCM wave function of the nucleus can be expressed as~\cite{saito1977PoTPS11}
\begin{equation}
    \Psi^\mathrm{GCM}=\int d\alpha f(\alpha)\Phi(\alpha),
\end{equation}
in which $d\alpha=d\alpha_1 d\alpha_2\cdots$, $\Phi(\alpha)$ is the generating wave function specified by the generator coordinates $\alpha_1, \alpha_2, \cdots$ serving as the variation parameters rather than physical coordinates.
The weight function $f(\alpha)$ is determined by solving the Hill-Wheeler equation
\begin{equation}
    \int \left[H(\alpha,\alpha')-EN(\alpha,\alpha')\right]f(\alpha')d\alpha'=0,
\end{equation}
in which the Hamiltonian and norm kernels are defined by
\begin{equation}
\begin{aligned}
    &H(\alpha,\alpha')=\left<\Phi(\alpha)\left|\mathcal{H}\right|\Phi(\alpha')\right> \cr
    &N(\alpha,\alpha')=\left<\Phi(\alpha)\middle|\Phi(\alpha')\right>.
\end{aligned}
\end{equation}

The GCM is extensively used in the analyses of clustering structures in nuclei, combined with the Brink wave function serving as the basis wave functions to be superposed to obtain the wave function of the total system.
The Brink wave function is defined as the fully antisymmetrized many-body wave function, consisting of several cluster wave functions characterized by various generator coordinates.
For a nucleus including $A$ nucleons with the clustering configuration of $C_1+C_2+\cdots+C_N$, the Brink wave function is given by~\cite{brink1966PotISoPEFVC3}
\begin{equation}
    \Phi^\mathrm{B}(\boldsymbol R_1,\cdots,\boldsymbol R_N)=\mathcal{A}\left\{\Phi_{C_1}(\boldsymbol{R}_1)\cdots\Phi_{C_N}(\boldsymbol{R}_N)\right\},
\end{equation}
in which $C_j$ ($j=1,\cdots,N$) denotes the $j$-th cluster and its mass number.
The wave function of cluster $C_j$ is defined as follows:
\begin{equation}
\Phi_{C_j}(\boldsymbol R_j)=\mathcal{A}\{\phi_1(\boldsymbol R_j)\cdots\phi_{C_j}(\boldsymbol R_j)\},
\end{equation}
and $\boldsymbol{R}_j$ is the generator coordinate.
The single-particle wave function is taken as the Gaussian form
\begin{equation}
    \phi_k(\boldsymbol R_j)=\frac{1}{(\pi b^2)^{3/4}}\exp\left[-\frac{1}{2b^2}(\boldsymbol{r}_k-\boldsymbol{R}_j)^2\right] \chi_k \tau_k
\end{equation}
with the spins and isospins are fixed as up $\left|\uparrow\right>$ or down $\left|\downarrow\right>$.
$b$ denotes the harmonic-oscillator parameter of single nucleons.

To describe the realistic nuclear system, the GCM-Brink wave function is defined as the superposition of the angular-momentum and parity-projected Brink wave functions
\begin{equation}
    \Psi^{J\pi}_M = \sum_{i,K}c_{i,K} \hat{P}_{MK}^{J\pi}\Phi^\mathrm{B}(\{\boldsymbol R\}_i),
\end{equation}
with the projectors defined as
\begin{equation}
\begin{aligned}
    \hat{P}^J_{MK}&=\frac{2J+1}{8\pi^2}\int d\Omega D^{J*}_{MK}(\Omega)\hat{R}(\Omega) \cr
    \hat{P}^\pi&=\frac{1+\pi\hat{P}_r}2, \quad \pi=\pm,
\end{aligned}
\end{equation}
and index $i$ specifying each cluster configuration indicated by a set of generator coordinates $\{\boldsymbol R\}=\{\boldsymbol R_1,\cdots,\boldsymbol R_N\}$. 
Then, the coefficients $\{c_{i,K}\}$  are determined by solving the Hill-Wheeler equation. 

In the Brink cluster model, the existence of clusters is assumed \textit{a priori}.
In contrast, more flexible theoretical methods for studying nuclear clustering, such as antisymmetrized molecular dynamics (AMD)~\cite{ono1992PRL2898,ono1992PTP1185,kanada-enyo1995PRC628} and fermionic molecular dynamics (FMD)~\cite{feldmeier1990NPA147,feldmeier1995NPA493,feldmeier2000RMP655}, treat all nucleons independently without assuming any clustering structure. 
The Brink wave function can be considered as a special case of AMD or FMD wave functions, with a fixed harmonic-oscillator parameter $b$ and frozen degrees of freedom for the nucleons in clusters.

\subsection{Orthogonality condition model}

In RGM and GCM, the forbidden states, which stem from the Pauli exclusion between fermions, will be eliminated in the process of solving the equations of motion, as a consequence of the antisymmetrization in the nuclear systems.
As a semi-microscopic method, the orthogonality condition model (OCM)~\cite{saito1968PoTP893,saito1969PoTP705,saito1977PoTPS11} reduces the computation cost by removing the forbidden states artificially before solving the equation of motion.
Considering the semi-microscopic approximation of the RGM, whose equation of motion can be expressed as
\begin{equation}
    (\mathcal{H}-E\mathcal{N})\chi=0,
\end{equation}
where $\mathcal{H}$ and $\mathcal{N}$ are the Hamiltonian and normalization operators, respectively.
Note that the RGM equation can be rewritten as
\begin{equation}
    \Lambda(\mathcal{N}^{-1/2}\mathcal{H}\mathcal{N}^{-1/2}-E)\Lambda(\mathcal{N}^{1/2}\chi)=0,
\end{equation}
in which $\Lambda\equiv1-\sum_F\left|\chi_F\right>\left<\chi_F\right|$ with $\chi_F$ denoting the forbidden states.
In OCM, we assume that the non-locality effect of the RGM can be approximated by the orthogonality operator $\Lambda$ with respect to the Pauli-forbidden space. 
As a result, before the operation of orthogonality, $\mathcal{N}^{-1/2}\mathcal{H}\mathcal{N}^{-1/2}$ is approximated firstly by the Hamiltonian without the exchange part, e.g., with an effective localized potential $V^\mathrm{eff}$
\begin{equation}
    \mathcal{N}^{-1/2}\mathcal{H}\mathcal{N}^{-1/2}\approx -\frac{\hbar^2}{2\mu}\nabla_r^2+V^\mathrm{eff}(\boldsymbol r).
\end{equation}
In this case, we obtain the OCM equation
\begin{equation}
    \Lambda\left[-\frac{\hbar^2}{2\mu}\nabla_r^2+V^\mathrm{eff}(\boldsymbol r)-E\right]\Lambda\left(\mathcal{N}^{1/2}\chi\right)=0,
\end{equation}
which is much easier to solve even for heavier nuclei.

\subsection{Tohsaki-Horiuchi-Schuck-R\"opke wave function} 

In 2001, the Tohsaki-Horiuchi-Schuck-R\"opke~(THSR) wave function~\cite{tohsaki2001PRL192501} was originally proposed to describe the $\alpha$-condensation of the Hoyle state, defined as
\begin{equation}
\begin{aligned}
    &\Psi^{\mathrm{THSR}}_{3\alpha} (B) \cr 
    &\quad=N(B) \mathcal{A}\left\{\exp\left[-\frac{2}{B^2}\sum_{k=1}^3(\boldsymbol{X}_k-\boldsymbol{X}_\mathrm{cm})^2\right]\prod_{i=1}^3\phi(\alpha_i)\right\}
\end{aligned}
\end{equation}
where $B=(b^2+2R_0^2)^{1/2}$ and $\boldsymbol{X}_i$ denotes the coordinate of c.o.m.\ of the $i$-th $\alpha$ cluster, whose internal wave function is $\phi(\alpha_i)$.
It can be seen that, in the THSR wave function, the 3 $\alpha$-clusters are occupying the same lowest $0S$ harmonic-oscillator orbit characterized by a width parameter $B$.
When the $B$ parameter is as large as the size of the whole nucleus, the $\alpha$ clusters can be considered to be moving freely in the nucleus, occupying the lowest $0S$ orbit.
Such a behavior of the $\alpha$ clusters is associated with the concept of Bose-Einstein condensation~(BEC) of bosons. 
On the other hand, in the limit when $B$ has the same value of the width parameter of the free $\alpha$-particle, $B=b$, the THSR wave function is reduced to the Brink wave function.

An important feature of clustering revealed by the THSR wave function is the BEC nature of the Hoyle state.
It was found that, by variation of the total energy, the Hoyle state can be obtained using the THSR wave function with a rather large $B$ value, which means that the Hoyle state can be well interpreted as the 3-$\alpha$ condensate state, where the $\alpha$ clusters are all moving in the $0S$ orbit within a relatively large volume, consisting with the large radius of the Hoyle state.
More interestingly, a subsequent study showed that the THSR wave function of the Hoyle state is almost equivalent to the $3\alpha$ cluster-model wave functions obtained from RGM or GCM~\cite{funaki2003PRC051306}, e.g.,
\begin{equation}
    \left|\left<\Psi^\mathrm{THSR}_{3\alpha}\middle|\Psi^\mathrm{RGM/GCM}_{3\alpha}\right>\right|^2\approx100\%.
\end{equation}


Starting from the non-localized character of cluster motion, the container picture was proposed~\cite{zhou2014PRC034319,zhou2014PTEP101D01,zhou2022FS26} in which the size parameters $\{B_i\}$ of cluster relative-motion wave functions are considered as true dynamical quantities for describing the correlations between clusters and the cluster correlations are also considered. 
Furthermore, by addressing the separation of the center-of-mass problem and considering different cluster correlations, a new trial wave function was proposed~\cite{zhou2018PTEP041D01}, 
\begin{widetext}
\begin{equation}
\begin{aligned}
    \Psi^\mathrm{new}
    ={}&\hat{L}_{N-1}(\boldsymbol\beta)\hat{G}_N(\beta_0)\hat{D}(\boldsymbol{Z})\Phi_0(\boldsymbol{r}) \cr
    ={}&\int d^3\tilde{T}_1\cdots d^3\tilde{T}_{n-1} \exp\left[-\sum_{i=1}^{n-1}\frac{\tilde{T}_i^2}{\beta_i^2}\right]
    \int d^3R_1\cdots d^3R_n \exp\left[-\sum_{i=1}^n \frac{C_i(\boldsymbol{R}_i-\boldsymbol{Z}_i-\boldsymbol{T}_i)^2}{\beta_0^2-2b_i^2}\right]
    \Phi_0(\boldsymbol{r}-\boldsymbol{R}) \cr 
    ={}&n_0\exp\left[-\frac{A}{\beta_0^2}X_\mathrm{cm}^2\right] \mathcal{A}\left\{\prod_{i=1}^{N-1}\exp\left[-\frac{(\boldsymbol\xi_i-\boldsymbol{S}_i)^2}{2B_i^2}\right]\prod_{i=1}^N\phi_i^\mathrm{int}(b_i)\right\},
\end{aligned}
\label{eq:IGM_wf}
\end{equation}    
\end{widetext}
in which $\hat{D}(\boldsymbol{Z})$ shifts the nucleons to the positions of corresponding clusters, $\hat{G}_N(\beta_0)$ performs an integral transformation for separating the c.o.m.\ part from the wave function, and $\hat{L}_{N-1}(\boldsymbol\beta)$ describes the cluster correlations in the spirit of the container picture.
For more details, see Ref.~\cite{zhou2018PTEP041D01}.
The physical meanings of quantities in Eq.~(\ref{eq:IGM_wf}) are clear.
$\boldsymbol{X}_\mathrm{cm}$ is the c.o.m.\ coordinate of the total nuclear.
$\boldsymbol{\xi}_i$ and $\boldsymbol{S}_i$ are the Jacobi coordinates of the cluster c.o.m.\ coordinate $\boldsymbol{X}_i$ and the generator coordinate $\boldsymbol{Z}_i$, respectively.
The internal wave function of the $i$-th cluster $\phi^\mathrm{int}_i(b_i)$ depends on the width variable $b_i$ and includes the spin and isospin parts.
By performing the integral we can find the width of the Gaussian relative wave function
\begin{equation}
    B_k^2=\frac12\left[\sum_{i=1}^{k+1}C_i/\left(C_{k+1}\sum_{i=1}^k C_i\right)\right]\beta_0^2+\frac12\beta_k^2.
\end{equation}
In the future, by utilizing this wave function, we can achieve a more realistic description of various cluster states in light nuclei.

\section{Reduced-width amplitudes}
\label{rwa}

\begin{figure*}[!tb]
\includegraphics
  [width=0.8\hsize]
  {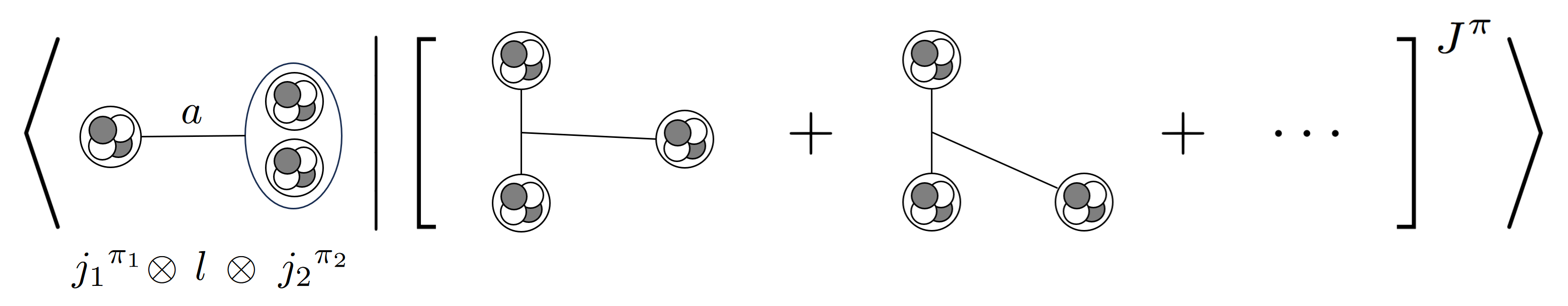}
\caption{The illustration of RWA taking the $\alpha+{}^8\mathrm{Be}$ structure in $^{12}\mathrm{C}$ as an example. On the left of the bracket is the channel wave function where the two clusters, with the angular momenta and parities of ${j_1}^{\pi_1}$ and ${j_2}^{\pi_2}$ respectively, are coupled together with the orbital angular momentum $l$. On the right is the GCM wave function, which superposes many clustering configurations and is projected on the angular momentum and parity $J^\pi$.}
\label{fig:rwa_coup}
\end{figure*}

Provided the microscopic wave functions based on the aforementioned cluster models, namely $\Psi^{J\pi}_M$, the reduced-width amplitude (RWA) of a system with the clustering structure of $A=C_1+C_2$ is defined as follows: 
\begin{equation}
\begin{aligned}
    &y_c^{J\pi}(a)=\sqrt{\frac{A!}{(1+\delta_{C_1C_2})C_1!C_2!}}  \cr
    &\quad\times\left<\frac{\delta(r-a)}{r^2}\left[Y_l(\hat{\boldsymbol r})\otimes
    \left[\Phi_{C_1}^{j_1\pi_1}\otimes\Phi_{C_2}^{j_2\pi_2}\right]_{j_{12}}\right]_{JM}\middle|\Psi^{J\pi}_M\right> ,
\end{aligned}
\end{equation}
in which $a$ is the distance between the clusters, and $c=\{j_1\pi_1 j_2\pi_2 j_{12}l\}$ denotes the coupling channel $[[C_1(j_1)\otimes C_2(j_2)]_{j_{12}}\otimes l]_J$, meaning that the clusters $C_1$ and $C_2$, with the angular momenta of $j_1$ and $j_2$ respectively, are coupled to the angular momentum of $j_{12}$, and then coupled with the orbital angular momentum $l$ to the total angular momentum of the nucleus $J$.
For each channel, the parities of the nucleus and clusters satisfy $\pi=(-)^l\pi_1\pi_2$.
$\Phi_{C_1}^{j_1\pi_1}$ and $\Phi_{C_2}^{j_2\pi_2}$ are reference wave functions of the clusters $C_1$ and $C_2$, respectively.
The $(1+\delta_{C_1C_2})$ factor comes from the exchange symmetry of the identical clusters $C_1=C_2$, and will be omitted in the following discussions. 
For clarity, we illustrate the above definition of RWA and the coupling channel in Fig.~\ref{fig:rwa_coup}.

The significance of RWA is twofold. 
Firstly, the RWA provides important information about the clustering configurations as well as the angular-momentum-coupling channels.
The amplitudes are directly related to the probabilities of forming a cluster structure at different separation distances.
As a result, the optimized distance of two-body clustering and the forbidden states can be inferred from the amplitudes.
To further evaluate the components of clustering configurations in a state of a nucleus, we can calculate the spectroscopic factors~(SF) by integrating the squared norm of RWA: 
\begin{equation}
    S_c^2=\int_0^\infty \left|a y_c^{J\pi}(a)\right|^2 da .
\label{s_factor}
\end{equation}
For narrow resonance states, according to the R-matrix theory~\cite{descouvemont2010RPP036301}, the RWA and SF are important parameters to deduce the decay parameters. 
The decay width can be calculated by
\begin{equation}
    \Gamma_c=2P_l(a)\gamma_c^2(a),
\end{equation}
with
\begin{equation}
    P_l(a)=\frac{ka}{F_l^2(ka)+G_l^2(ka)},
\end{equation}
in which $k$ is the momentum of inter-cluster motion in the asymptotic region, and $F_l(r)$ and $G_l(r)$ are the regular and irregular Coulomb functions, respectively.
$\gamma_c^2(a)$ is the reduced width, which can be approximated by the value of RWA
\begin{equation}
    \gamma_c^2(a)=\frac{\hbar^2}{2\mu a}\left[ay_c^{J\pi}(a)\right]^2,
\end{equation}
with $\mu$ denoting the reduced mass of two clusters.
As a good measure of the cluster-formation probability at the nuclear surface, the dimensionless reduced width is defined as the ratio of the reduced width to its Wigner limit $\gamma_W(a)=3\hbar^2/2\mu a$
\begin{equation}
    \theta_c^2(a)\equiv\frac{\gamma_c^2(a)}{\gamma_W^2(a)}=\frac{a}3\left[ay_c^{J\pi}(a)\right]^2.
\end{equation}

Another important aspect of the RWA lies in its application to reaction theories, serving as the input parameters for calculating the cross sections.
The reduced-width amplitude has essentially the equivalent physical meaning as the overlap function, which is usually employed in the reaction theories to evaluate the cross sections, defined as~\cite{timofeyuk2014JPGNPP094008}
\begin{equation}
\begin{aligned}
    I_{J_BJ_C}^{J_A}(\boldsymbol r)
    &=\sqrt{\frac{A!}{B!C!}}\left<\Psi_B^{J_B}\Psi_C^{J_C}\middle|\Psi_A^{J\pi}\right>  \cr
    &=\sum_{j_{12}m_{12}lm}I_c^{J\pi}(r)Y_{lm}(\hat{\boldsymbol r}) 
\end{aligned}
\end{equation}
for a reaction process where $A=B+C$.
The RWA, or overlap function, connecting the states in entrance and exit channels, is an important constituent in the scattering or transition matrix.
For example, in a $(d,p)$ transfer reaction, the scattering matrix from the initial state $i$ to a final state $f$ is~\cite{descouvemont2022EPJA193,chien2023PRC044605}
\begin{equation}
\begin{aligned}
    U^{J\pi}_{i,f}&=-\frac{i}{\hbar}\left<\Psi_f^{J\pi}\left| V_{pn}+\Delta V\right|\Psi_i^{J\pi}\right> \cr 
    &=-\frac{i}{\hbar}\sum_\gamma \int u_\gamma^{J\pi}(R)K_\gamma^{J\pi}(R,R')u_f^{J\pi}(R')RR' dR dR'
\end{aligned}
\end{equation}
in which $\Psi^{J\pi}_i$ and $\Psi^{J\pi}_f$ are the wave functions of the initial and final states, respectively, and $\Delta V$ is the remnant potential.
The function $u_\gamma^{J\pi}(R)$ is the radial function between the target nucleus and the deuteron in the reaction channel $\gamma$, and $u_f^{J\pi}$ is the radial function between the residual nucleus and the proton.
The transfer kernel is evaluated with the help of the RWA $y_c(r')$: 
\begin{equation}
\begin{aligned}
    K_\gamma^{J\pi}(R,R')={}&\mathcal{J}\sum_c\left<\left[\phi_k^l(\boldsymbol{r})\otimes Y_{L_i}(\Omega_R)\right]^J\right|V_{pn} \cr 
    &+\Delta V\left|\left[y_c(r')\otimes Y_{L_f}(\Omega_{R'})\right]^J\right>,
\end{aligned}
\end{equation}
where $\mathcal{J}$ is the Jacobian, $\phi_k^{lm}$ is the CDCC (continuum discretized coupled channel) wave function of deuteron.

On the other hand, for $(p,p\alpha)$ knock-out reactions, the triple-differential cross section (TDX) can be evaluated using the distort-wave impulsive approximation (DWIA)~\cite{yoshida2019PRC044601,yoshida2018PRC024614}
\begin{equation}
    \frac{d^3\sigma}{dE_1^\mathrm{L} d\Omega_1^\mathrm{L} d\Omega_2^\mathrm{L}}=F_\mathrm{kin}C_0\frac{d\sigma_{p\alpha}}{d\Omega_{p\alpha}}(\theta_{p\alpha},T_{p\alpha})\left|\bar{T}_{\boldsymbol{K}_i}\right|^2,
\end{equation}
in which the kinematical factor $F_\mathrm{kin}$ is defined as
\begin{equation}
    F_\mathrm{kin}=J_\mathrm{L}\frac{K_1 K_\alpha E_1 E_\alpha}{(\hbar c)^4}\left[1+\frac{E_\alpha}{E_\mathrm{B}}+\frac{E_\alpha}{E_\mathrm{B}}\frac{\boldsymbol{K}_1\cdot\boldsymbol{K}_\alpha}{K_\alpha^2}\right]
\end{equation}
with $J_\mathrm{L}$ denoting the Jacobian form the c.o.m\ frame to the L frame, and 
\begin{equation}
    C_0=\frac{E_0}{(\hbar c)^2K_0}\frac{\hbar^4}{(2\pi)^3\mu_{p\alpha}^2}
\end{equation}
is a constant.
$\boldsymbol{K}_i$, $\Omega_i$, and $E_i$ denote the wave number and its solid angle, and the total energy of particle $i$ ($i=0$ for incident proton, $=1$ for emitted proton, and $=\alpha$ for emitted $\alpha$ cluster), respectively.
The reduced transition matrix is related to the $\alpha$-RWA by
\begin{equation}
    \bar{T}_{\boldsymbol{K}_i}=\int d\boldsymbol{r} F_{\boldsymbol{K}_i}(\boldsymbol{r})y(r)Y_{00}(\hat{\boldsymbol{r}}),
\end{equation}
where
\begin{equation}
    F_{\boldsymbol{K}_i}(\boldsymbol{r})=\chi_{1,\boldsymbol{K}_1}^{*(-)}(\boldsymbol{r})\chi_{\alpha,\boldsymbol{K}_\alpha}^{*(-)}(\boldsymbol{r})\chi_{0,\boldsymbol{K}_0}^{(+)}(\boldsymbol{r})e^{-i\boldsymbol{K}_0\cdot\boldsymbol{r}A_\alpha/A}.
\end{equation}
$\chi_{i,\boldsymbol{K}_i}$ is the distorted wave between the particle $i$ and $A$ for $i=0$, and between $i$ and $B$ otherwise.
The superscripts $(+)$ and $(-)$ indicate the outgoing and incoming boundary conditions of the scattering waves, respectively.

\subsection{Calculation methods of RWA}

The calculation methods of RWA have been established for many years~\cite{horiuchi1977PoTPS90}.
In RGM, the calculation of RWA  is straightforward since the RGM-type wave function can be written as
\begin{equation}
\begin{aligned}
    &\Psi^\mathrm{RGM}=\sqrt{\frac{C_1!C_2!}{A!}} \cr
    &\qquad\times\mathcal{A}\left\{\chi_l(r)\left[Y_l(\hat r)\otimes\left[\Phi_{C_1}^{j_1\pi_1}\otimes\Phi_{C_2}^{j_2\pi_2}\right]_{j_{12}}\right]_{JM}\right\}.
\end{aligned}
\label{rgm_wf2}
\end{equation}
The relative-motion wave function $\chi_l(r)$ can be expanded using radial harmonic oscillator functions $R_{nl}(r,\nu')$
\begin{equation}
    \chi_l(r)=\sum_n e_n R_{nl}(r,\nu'),
\label{chi_l}
\end{equation}
in which
\begin{equation}
    e_n=\int R_{nl}(r,\nu') \chi_l(r) r^2 dr.
\end{equation}
and $\nu'=(C_1 C_2/A)\nu$ with $\nu=1/2b^2$ denoting the width parameter. 
Then, the RWA can be calculated as
\begin{equation}
    y_c^{J\pi}(r)=\sum_n \mu_{nl} e_n R_{nl}(r,\nu')
\label{rwa_rgm}
\end{equation}
where $\mu_{nl}$ are the eigenvalues of the RGM norm kernel
\begin{widetext}
\begin{equation}
    \mu_{nl}=\left<R_{nl}(r)\left[Y_l(\hat r)\left[\Phi_{C_1}^{j_1\pi_1}\Phi_{C_2}^{j_2\pi_2}\right]_{j_{12}}\right]_J\middle|\mathcal{A}\left\{R_{nl}(r)\left[Y_l(\hat r)\left[\Phi_{C_1}^{j_1\pi_1}\Phi_{C_2}^{j_2\pi_2}\right]_{j_{12}}\right]_J\right\}\right> .
\label{rgm_nor}
\end{equation}
Within the framework of GCM-Brink, the coefficients $e_{nl}$ can be calculated as~\cite{horiuchi1977PoTPS90}
\begin{equation}
    e_{nl}=(-)^{(n-l)/2}\sqrt{\frac{2l+1}{(n-l)!!(n+l+1)!!}}\sum_{pq}\frac{(\nu S_p^2)^{n/2}}{\sqrt{n!}} e^{-\nu S_p^2/2}B_{pq}^{-1}\left<\Phi_{j_1\pi_1j_2\pi_2j_{12}l}^{J\pi}(S_q)\middle|\Psi^{J\pi}_{MA}\right>, 
\end{equation}
where
\begin{equation}
    B_{pq}=\left<\Phi_{j_1\pi_1j_2\pi_2j_{12}l}^{J\pi}(S_p)\middle|\Phi_{j_1\pi_1j_2\pi_2j_{12}l}^{J\pi}(S_q)\right>,
\end{equation}
$\Phi_{j_1\pi_1j_2\pi_2j_{12}l}^{J\pi}(S_p)$ is the Brink wave function projected onto the angular momenta and parities of the nucleus and two clusters involved, and $S_p$ is the discretized inter-cluster distance.

The traditional method for calculating the RWA using the GCM-Brink wave function requires enormous computational effort.
Recently, Chiba and Kimura~\cite{chiba2017PTEP053D01} proposed the Laplace expansion method for calculating the RWA within the GCM/AMD framework.
Using Laplace expansion, the AMD wave function of the $A$-nucleon system, defined as the determinant of an $A\times A$ matrix $B_{A\times A}$, can be split into two AMD wave functions of clusters $C_1$ and $C_2$ ($A=C_1+C_2$)
\begin{equation}
    \Phi_A^\mathrm{AMD}=\sqrt{\frac{C_1!C_2!}{A!}}\sum_{1\leq i_1<\cdots<i_{C_1}\leq A} P(i_1,\cdots,i_{C_1}) 
    \Phi_{C_1}^\mathrm{AMD}(i_1,\cdots,i_{C_1})\Phi_{C_2}^\mathrm{AMD}(i_{C_1+1},\cdots,i_A)
\end{equation}
with the phase factor $P(i_1,\cdots,i_{C_1})$ defined as 
\begin{equation}
    P(i_1,\cdots,i_{C_1})=(-)^{C_1(C_1+1)/2+\sum_{s=1}^{C_1}i_s}.
\end{equation}
The cluster wave function $\Phi_{C_1}^\mathrm{AMD}(i_1,\cdots,i_{C_1})$ is defined as the determinant of the matrix composed of the $1,\cdots,C_1$th rows and the $i_1,\cdots,i_{C_1}$th columns of the matrix $B_{A\times A}$, and the determinant defining  $\Phi_{C_2}^\mathrm{AMD}(i_{C_1+1},\cdots,i_A)$ consists of the elements left after removing the $1,\cdots,C_1$th rows and the $i_1,\cdots,i_{C_1}$th columns.
Then, the RWA can be calculated as the summation of the coupling results of three kernels with angular-momentum and parity
\begin{equation}
\begin{aligned}
    y_c^{J\pi}(a)=\frac1{\sqrt{\mathcal{N}^{J\pi}_K}}\sum_{1\leq i_1<\cdots<i_{C_1}\leq A}&P(i_1,\cdots,i_{C_1}) \cr
    &\times\left[\chi_l(a;i_1,\cdots,i_A)\otimes\left[N^{j_1\pi_1}(i_1,\cdots,i_{C_1})\otimes N^{j_2\pi_2}(i_{C_1+1},\cdots,i_A)\right]_{j_{12}}\right]_{JK} ,
\end{aligned}
\end{equation}
in which
\begin{equation}
\begin{aligned}
    \chi_{lm_l}(a;i_1,\cdots,i_A)&=\left<\frac{\delta(r-a)}{r^2}Y_{lm_l}(\hat{r})\middle|\chi(\boldsymbol{r};i_1,\cdots,i_A)\right>  \\
    N^{j_1\pi_1}_{m_1}(i_1,\cdots,i_{C_1})&=\left<\Phi^{j_1\pi_1}_{m_1C_1}\middle|\Phi^\mathrm{int}_{C_1}(i_1,\cdots,i_{C_1})\right> \\
    N^{j_2\pi_2}_{m_2}(i_{C_1+1},\cdots,i_A)&=\left<\Phi^{j_2\pi_2}_{m_2C_2}\middle|\Phi^\mathrm{int}_{C_2}(i_{C_1+1},\cdots,i_A)\right>.
\end{aligned}
\end{equation}
\end{widetext}
It should be noted that the application of the Laplace expansion method on symmetric clustering structure is time-consuming, as the number of possible combinations for Laplace expansion becomes huge when the system is heavy and the cluster mass number $C_1$ is close to $C_2$.
In that case, the traditional method of calculation can be applied more efficiently.

\subsection{Asymptotic behavior}

\begin{figure*}[!tb]
\includegraphics
  [width=0.9\hsize]
  {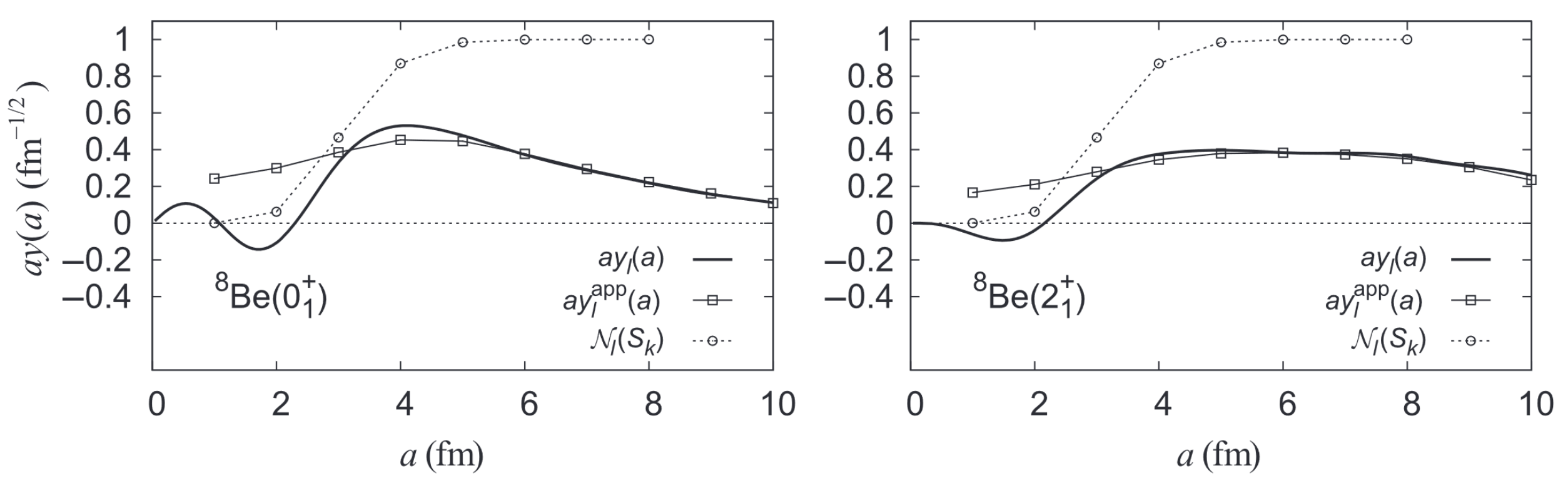}
\caption{The RWA and approximated RWA of $\alpha+\alpha$ channel in $0^+_1$ and $2^+_1$ states of $^8\mathrm{Be}$. 
The Figure is taken from Ref.~\cite{kanada-enyo2014PTEP073D02}.}
\label{fig:rwa_Be8}
\end{figure*}

For cluster states in self-conjugate nuclei, such as $\alpha+\alpha$ as shown in Fig.~\ref{fig:rwa_Be8}, the RWA exhibits distinct features in different regions, namely the suppressed inner oscillation, the enhanced surface peak, and the damping outer tail.
The inner oscillation and the enhanced peak are intimately related to the antisymmetrization effect between clusters.
Due to the fermionic nature of the nucleons in the $\alpha$ clusters, the formation probability of $\alpha+\alpha$ structure at a small distance is suppressed and forbidden states appear at the nodal distances.
On the other hand, the tail part of RWA, where the distance between clusters is large enough that the antisymmetrization effect between clusters becomes very weak, is mainly determined by the separation energy as well as the centrifugal and Coulomb barriers.
Consequently, the asymptotic behavior of RWA should be well-defined and is important for discussing the delocalization of the $\alpha$ clusters in weakly bound cluster states~\cite{kanada-enyo2014PoTaEP103D03}.
Furthermore, in analyzing nuclear reactions, the asymptotic behavior of RWA (overlap function) determines the angular distributions of nucleon or cluster removal cross sections~\cite{timofeyuk2014JPGNPP094008}.
For bound systems and narrow resonances, the tail part of RWA decreases as
\begin{equation}
    ay_c^{J\pi}(a) \to C_c^{J\pi}W_{-\eta,l+1/2}(2\kappa a),
\end{equation}
in which $W_{-\eta,l+1/2}(2\kappa a)$ is the Whittaker function, $\eta=Z_1Z_2e^2\mu/\hbar^2\kappa$ is the Sommerfeld parameter, and $\kappa=\sqrt{-2\mu E}/\hbar$ is the wave number.  
$C_c^{J\pi}$ is the asymptotic normalization coefficient~(ANC) and plays many important roles in reaction analyses.
 
The significance of the asymptotic behavior of RWA is further demonstrated by the consistency between the tail parts of relative wave functions with distinctive definitions.
It should be noted that the RWA is not normalized to unity, but to $S_c^2$ as shown in Eq.~(\ref{s_factor}). 
If we define
\begin{equation}
    u_l(r)=\sum_n e_n\sqrt{\mu_{nl}}R_{nl}(r),
\label{u_l}
\end{equation}
it can be seen that, for the cluster-model wave function normalized to unity, e.g., $\left<\Psi|\Psi\right>=1$, we have
\begin{equation}
    \int |u_l(r)|^2 r^2dr =1.
\end{equation}
The difference between the relative wave functions $y_l(r)$, $u_l(r)$, and $\chi_l(r)$ lies in the treatment of the antisymmetrization effect between clusters~\cite{kanada-enyo2014PTEP073D02}.
By observing Eq.~(\ref{rgm_wf2}), we can see that $\chi_l(r)$ is the cluster relative wave function before antisymmetrization, and as a result, contains the nonphysical forbidden states whose norm kernel has the eigenvalue of $\mu_{nl}=0$.
On the contrary, in the antisymmetrized relative wave functions $y_l(r)$ and $u_l(r)$, the forbidden-state components are eliminated by the coefficient $\mu_{nl}$ and $\sqrt{\mu_{nl}}$, respectively, while the treatments of the partially allowed states are different and thus exhibit different normalization results.
Such distinctions can be inferred more explicitly by comparing Eq.~(\ref{rwa_rgm}), (\ref{u_l}), and (\ref{chi_l}).
Notably, although the wave functions $\chi_l(r)$, $u_l(r)$, and $y_l(r)$ are different in the region where antisymmetrization effect plays an important role, i.e., the inner region with a small distance between clusters, they have the same asymptotic behavior in the region with large $r$, where the antisymmetrization effect is weak and can be ignored, e.g., for large $r$
\begin{equation}
    \chi_l(r)\approx y_l(r) \approx u_l(r).
\end{equation}
Using this important feature, Kanada-En'yo et al.~\cite{kanada-enyo2014PTEP073D02} have shown that, for two-body cluster channels, the tail part of RWA can be well approximated using the overlap of the cluster-model wave function with the single-Brink wave function
\begin{equation}
    |ay_l(a)|\approx \frac1{\sqrt2}\left(\frac{2\nu'}{\pi}\right)^{1/4}\left|\left<\Psi^{J\pi}_M\middle|\Phi^\mathrm{B}(S=a)\right>\right|\equiv ay^\mathrm{app}(a),
\label{rwa_app}
\end{equation}
with $S$ indicating the inter-cluster distance of the Brink wave function.
In Fig.~\ref{fig:rwa_Be8}, we show the approximated RWA, compared with the exact one, of $0_1^+$ and $2_1^+$ states of $^{8}\mathrm{Be}$, respectively.
It can be seen that the approximated RWA can well describe the tail part, though the inner oscillation is absent.
As mentioned in the beginning of the present section, in analyzing the decay characters, only the RWA value in a relatively larger channel radius $a$ is required.
In that case, the approximation of RWA using single-Brink overlap provides a practical calculation method by effectively reducing the computational cost.

It is also interesting to test the abilities of different trial wave functions in describing the asymptotic behavior of cluster relative motion.
Based on the equivalence of the three kinds of relative wave function in the tail region, Kanada-En'yo~\cite{kanada-enyo2014PoTaEP103D03} further calculated the relative wave functions obtained from the Brink wave function, the spherical THSR (sTHSR) wave function, the deformed THSR (dTHSR) wave function, and a function with Yukawa tail (YT), and compared them with the exact solution obtained by GCM wave function.
The relative wave functions of these trail wave functions can be written as various types of Gaussians.
For the Brink wave function and sTHSR wave function, the inter-cluster wave function can be adopted as the shifted spherical Gaussian (ssG)
\begin{equation}
    \chi^\mathrm{ssG}(\boldsymbol{r})=\exp\left[-\frac{(\boldsymbol{r}-\boldsymbol{S})^2}{\sigma^2}\right]
\end{equation}
whose partial wave expansion reads
\begin{equation}
    \chi_l^\mathrm{ssG}(S,\sigma;r)\propto i_l\left(\frac{2Sr}{\sigma^2}\right)\exp\left(-\frac{r^2+S^2}{\sigma^2}\right),
\end{equation}
where $i_l(r)$ is the regular modified spherical Bessel function.
This function is controlled by two parameters, $S$ and $\sigma$.
When $\sigma$ is fixed as $\sigma=1/\sqrt{\nu'}=\sqrt{A/C_1C_2}b$, the relative wave function corresponds to the Brink wave function, while when $S\to0$, the limit equals to the relative wave function of the sTHSR wave function.  
For the dTHSR wave function, the relative wave function is described by the deformed Gaussian (dG) function around the origin.
Take the axially symmetric case for example, 
\begin{equation}
\begin{aligned}
    \chi^\mathrm{dG}(\sigma_\bot,\sigma_z;\boldsymbol{r})&\propto \exp\left(-\frac{x^2}{\sigma_\bot^2}-\frac{y^2}{\sigma_\bot^2}-\frac{z^2}{\sigma_z^2}\right) \cr
    &=\exp\left(-\frac{r^2}{\sigma_\bot^2}+\frac{r^2}{\Delta}\cos\theta^2\right)
\end{aligned}
\end{equation}
with
\begin{equation}
    \frac1\Delta\equiv\frac1{\sigma_\bot^2}-\frac1{\sigma_z^2}.
\end{equation}
The partial wave function can be calculated as, for even-$l$ wave
\begin{equation}
\begin{aligned}
    \chi_l^\mathrm{dG}(\sigma_\bot,\sigma_z;r)\propto{}&2\sqrt{(2l+1)\pi}\exp\left(-\frac{r^2}{\sigma_\bot^2}\right) \cr 
    &\times\int_0^1 P_l(t)\exp\left(\frac{r^2}{\Delta}t^2\right)dt ,
\end{aligned}
\end{equation}
and for odd-$l$ wave
\begin{equation}
\begin{aligned}
    \chi_l^\mathrm{dG}(\sigma_\bot,\sigma_z;r)\propto{}&2\sqrt{(2l+1)\pi}r\exp\left(-\frac{r^2}{\sigma_\bot^2}\right) \cr 
    &\times\int_0^1 P_l(t)t\exp\left(\frac{r^2}{\Delta}t^2\right)dt,
\end{aligned}
\end{equation}
where $P_l(t)$ is the Legendre polynomial.

\begin{figure*}[!tb]
\includegraphics
  [width=0.9\hsize]
  {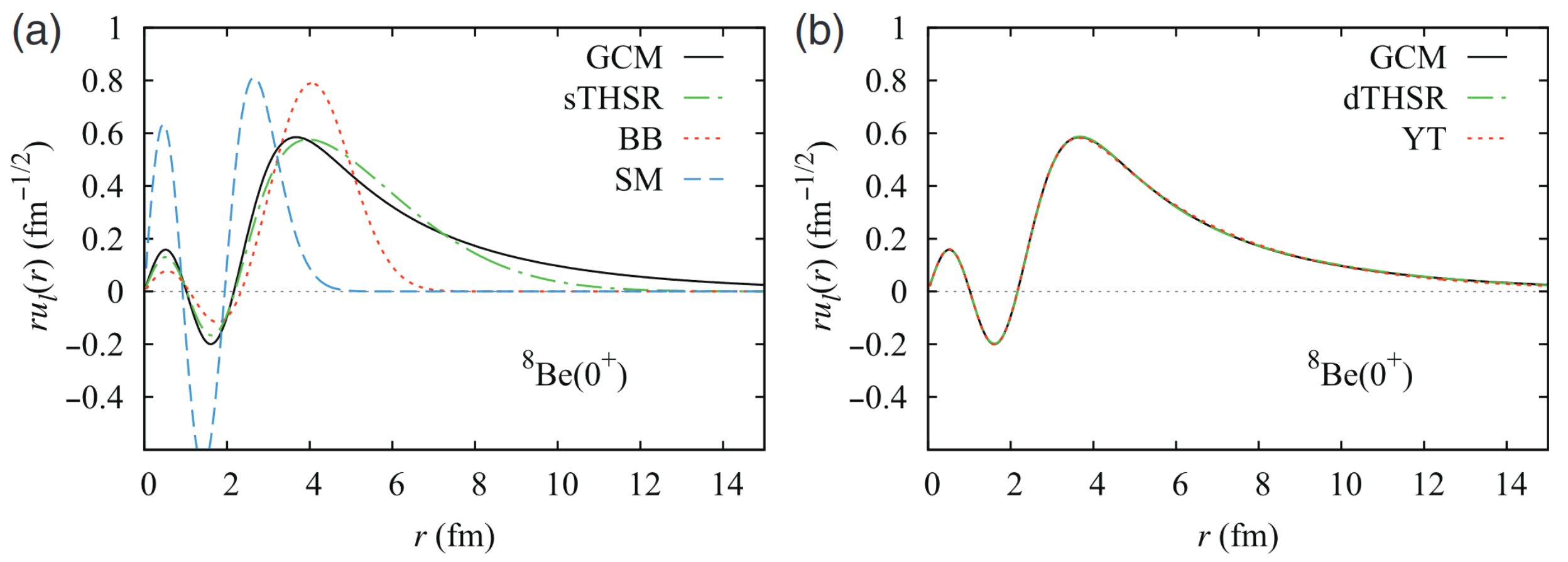}
\caption{The $\alpha$-$\alpha$ cluster relative wave function obtained by GCM wave function of $^8\mathrm{Be}(0^+)$, compared with that obtained by shell-model wave function, Brink wave function, spherical THSR wave function, deformed THSR wave function, and Yukawa-tail function. The Figure is taken from Ref.~\cite{kanada-enyo2014PoTaEP103D03}.}
\label{fig:rf_Be8}
\end{figure*}

To see the accuracy of the trial wave functions in describing the asymptotic behavior of cluster relative motions, we present in Fig.~\ref{fig:rf_Be8} the relative wave function $ru_l(r)$ obtained by precise GCM solutions and trial wave functions including the Brink wave function, spherical and deformed THSR wave functions, and Yukawa-type wave function calculated by Kanada-En'yo~\cite{kanada-enyo2014PoTaEP103D03}.
For comparison, the result of SM wave function is also displayed.
As the results show, the SM and Brink wave functions can only give the correct number of oscillation nodes in the inner region, but both decrease too fast in the tail region to give the reasonable asymptotic feature.
Besides, the SM wave function exhibits an enhanced peak narrower and more inward than the precise wave function, while the Brink wave function shows a more outward peak.
Notably, the Brink wave function reproduces better the amplitudes of the inner oscillation part and exhibits a longer tail compared to the SM wave function.
However, the tail of the Brink relative wave function deviates drastically from the exact one, as the Brink wave function is a localized model wave function.
By introducing the non-localization character, the sTHSR wave function can describe the relative wave function better compared to the Brink and SM wave functions, although minor deviations appear in the tail part.
The dTHSR wave function and the YT wave function can surprisingly give quite precise results of cluster relative wave function, suggesting that they can serve as the efficient trial wave function in describing the relative motion between clusters.

\subsection{Two-body overlap amplitude}
\label{subsec:3body_theor}

The RWA is essentially a one-body overlap amplitude that depends on a single inter-cluster distance parameter.
To more clearly observe the correlations between clusters or nucleons and to understand the much more complex three-body cluster motion, we can extend the analysis of overlap amplitudes to three-body channels.
Such extension has recently been applied in some works to analyze $\mathrm{core}+N+N$ structures~\cite{descouvemont2023PRC014312,kimura2017PRC034331,kobayashi2016PRC024310}.
Here, we attempt to provide a more general formula for calculating the two-body overlap amplitude by iteratively applying the Laplace expansion method~\cite{chiba2017PTEP053D01}.

The two-body overlap amplitude can be defined as
\begin{widetext}
\begin{equation}
\begin{aligned}
    \mathcal{Y}_c^{J\pi}(a_1,a_2)={}&\sqrt{\frac{A!}{C_1!C_2!C_3!}} \cr 
    &\times\left<\frac{\delta(r_1-a_1)\delta(r_2-a_2)}{r_1^2r_2^2}\left[\left[Y_{l_1}(\hat{r}_1)\otimes Y_{l_2}(\hat{r}_2)\right]_L\otimes\left[\Phi_{C_1}^{j_1\pi_1}\otimes\left[\Phi_{C_2}^{j_2\pi_2}\otimes\Phi_{C_3}^{j_3\pi_3}\right]_{j_{23}}\right]_{j_{123}}\right]_{JM}\middle|\Psi_{M}^{J\pi}\right> ,
\end{aligned}
\end{equation}
in which $c=\{j_1\pi_1j_2\pi_2j_3\pi_3j_{23}j_{123}l_1l_2L\}$ according to the three-body coupling channel of $[[C_1(j_1)\otimes[C_2(j_2)\otimes C_3(j_3)]_{j_{23}}]_{j_{123}}\otimes[l_1\otimes l_2]_L]_J$ and the parity relation of $\pi=(-)^{l_1+l_2}\pi_1\pi_2\pi_3$.
The relative-motion coordinates $\boldsymbol{r}_1$ and $\boldsymbol{r}_2$ are defined as 
\begin{equation}
\begin{aligned}
    \boldsymbol{r}_1&=\boldsymbol{X}_2-\boldsymbol{X}_3 \\
    \boldsymbol{r}_2&=\boldsymbol{X}_1-\frac{C_2\boldsymbol{X}_2+C_3\boldsymbol{X}_3}{C_2+C_3},
\end{aligned}
\label{coord_jacobi}
\end{equation}
where $\boldsymbol{X}_i$ is the c.o.m.\ of the physical coordinates of cluster $C_i$.
By applying the Laplace expansion method twice to the AMD/Brink wave function, which is defined as a Slater determinant, we have
\begin{equation}
\begin{aligned}
\Phi_A^\mathrm{AMD}=\sqrt{\frac{A!}{C_1!C_2!C_3!}}&\sum_{\substack{1\leq i_1<\cdots<i_{C_1}\leq A \\ 1\leq j_1<\cdots<j_{C_2}\leq C_2+C_3}} P_A(i_1,\cdots,i_{C_1})P_{C_2+C_3}(j_1,\cdots,j_{C_2}) \cr 
&\times\Phi_{C_1}^\mathrm{AMD}(i_1,\cdots,i_{C_1})\Phi_{C_2}^\mathrm{AMD}(j_1,\cdots,j_{C_2})\Phi_{C_3}^\mathrm{AMD}(j_{C_2+1},\cdots,j_{C_2+C_3})
\end{aligned}
\end{equation}
with the phase factor 
\begin{equation}
\begin{aligned}
P_A(i_1,\cdots,i_{C_1})&=(-)^{\frac{C_1(C_1+1)}2+\sum_si_s} \\
P_{C_2+C_3}(j_1,\cdots,j_{C_2})&=(-)^{\frac{C_2(C_2+1)}2+\sum_sj_s}.
\end{aligned}
\end{equation}
Note that $j_s$ $(s=1,\cdots,C_2+C_3)$ corresponds to the index in the matrix composed of the elements left after removing the rows and columns constituting the $\Phi_{C_1}^\mathrm{AMD}$ from the matrix $B_{A\times A}$. 
Following the analogues procedure in the two-body Laplace expansion method calculation, the product of three AMD wave functions can be rewritten as 
\begin{equation}
\begin{aligned}
\Phi_{C_1}^\mathrm{AMD}&\Phi_{C_2}^\mathrm{AMD}\Phi_{C_3}^\mathrm{AMD} \cr
&=\Phi_{C_1}^\mathrm{cm}\Phi_{C_2}^\mathrm{cm}\Phi_{C_3}^\mathrm{cm}
\Phi_{C_1}^\mathrm{int}\Phi_{C_2}^\mathrm{int}\Phi_{C_3}^\mathrm{int} \cr 
&=\left(\frac{C_1C_2C_3}{(\pi b^2)^3}\right)^{3/4}
\exp\left\{-\frac1{2b^2}\left[C_1(\boldsymbol{X}_1-\boldsymbol{Z}_1)^2+C_2(\boldsymbol{X}_2-\boldsymbol{Z}_2)^2+C_3(\boldsymbol{X}_3-\boldsymbol{Z}_3)^2\right]\right\}\Phi_{C_1}^\mathrm{int}\Phi_{C_2}^\mathrm{int}\Phi_{C_3}^\mathrm{int} \cr
&=\left(\frac{C_1C_2C_3}{(\pi b^2)^3}\right)^{3/4}
\exp\left\{-\frac1{2b^2}\left[AX_G^2+\frac{C_2C_3}{C_2+C_3}(\boldsymbol{r}_1-\boldsymbol{S}_1)^2+\frac{C_1(C_2+C_3)}{C_1+C_2+C_3}(\boldsymbol{r}_2-\boldsymbol{S}_2)^2\right]\right\}\Phi_{C_1}^\mathrm{int}\Phi_{C_2}^\mathrm{int}\Phi_{C_3}^\mathrm{int} \cr 
&=\Phi_A^\mathrm{cm}\chi_{23}(\boldsymbol{r}_1)\chi_{123}(\boldsymbol{r}_2)  \Phi_{C_1}^\mathrm{int}\Phi_{C_2}^\mathrm{int}\Phi_{C_3}^\mathrm{int} ,
\end{aligned}
\end{equation}
in which $\boldsymbol{Z}_i$ is the c.o.m.\ of the generator coordinates of nucleons in cluster $C_i$, and $\boldsymbol{S}_i$ is the corresponding Jacobi coordinate, 
\begin{equation}
    \Phi_A^\mathrm{cm}=\left(\frac{A}{\pi b^2}\right)^{3/4}\exp\left[-\frac{A}{2 b^2}X_G^2\right]
\end{equation}
is the c.o.m.\ wave function of the nucleus, and
\begin{equation}
\begin{aligned}
    \chi_{23}(\boldsymbol{r}_1)&=\left(\frac{C_2C_3}{C_2+C_3}\frac1{\pi b^2}\right)^{3/4}\exp\left[-\frac{C_2C_3}{C_2+C_3}\frac1{2 b^2}(\boldsymbol{r}_1-\boldsymbol{S}_1)^2\right] \\
    \chi_{123}(\boldsymbol{r}_2)&=\left(\frac{C_1(C_2+C_3)}{A}\frac1{\pi b^2}\right)^{3/4}\exp\left[-\frac{C_1(C_2+C_3)}{A}\frac1{2 b^2}(\boldsymbol{r}_2-\boldsymbol{S}_2)^2\right]
\end{aligned}
\end{equation}
are the relative wave functions between $C_2$ and $C_3$ and between $C_1$ and the c.o.m.\ of $C_2+C_3$, respectively. 
$\Phi_{C_1}^\mathrm{int}$, $\Phi_{C_2}^\mathrm{int}$, and $\Phi_{C_3}^\mathrm{int}$ are the intrinsic wave functions of $C_1$, $C_2$, and $C_3$.
Then, the two-body overlap amplitude can be calculated by
\begin{equation}
\begin{aligned}
&\mathcal{Y}_c^{J\pi}(a_1,a_2) \cr
={}&\sqrt{\frac{A!}{C_1!C_2!C_3!}}\left<\frac{\delta(r_1-a_1)\delta(r_2-a_2)}{r_1^2r_2^2}
\left[\left[Y_{l_1}(\hat{r}_1)\otimes Y_{l_2}(\hat{r}_2)\right]_L\otimes\left[\Phi_{C_1}^{j_1\pi_1}\otimes\left[\Phi_{C_2}^{j_2\pi_2}\otimes\Phi_{C_3}^{j_3\pi_3}\right]_{j_{23}}\right]_{j_{123}}\right]_{JM}\middle|\Psi_{M}^{J\pi}\right> \cr 
={}&\sqrt{\frac{A!}{C_1!C_2!C_3!}}\left<\frac{\delta(r_1-a_1)\delta(r_2-a_2)}{r_1^2r_2^2}P^{J\pi}_{KM}
\left[\left[Y_{l_1}(\hat{r}_1)\otimes Y_{l_2}(\hat{r}_2)\right]_L\otimes\left[\Phi_{C_1}^{j_1\pi_1}\otimes\left[\Phi_{C_2}^{j_2\pi_2}\otimes\Phi_{C_3}^{j_3\pi_3}\right]_{j_{23}}\right]_{j_{123}}\right]_{JM}\middle|\Psi_{A}^\mathrm{int}\right> \cr 
={}&\sqrt{\frac{A!}{C_1!C_2!C_3!}}\left<\frac{\delta(r_1-a_1)\delta(r_2-a_2)}{r_1^2r_2^2}
\left[\left[Y_{l_1}(\hat{r}_1)\otimes Y_{l_2}(\hat{r}_2)\right]_L\otimes\left[\Phi_{C_1}^{j_1\pi_1}\otimes\left[\Phi_{C_2}^{j_2\pi_2}\otimes\Phi_{C_3}^{j_3\pi_3}\right]_{j_{23}}\right]_{j_{123}}\right]_{JK}\middle|\Psi_{A}^\mathrm{int}\right> \cr 
={}&\frac1{\sqrt{\mathcal{N}^{J\pi}_K}}\sum_{\substack{1\leq i_1<\cdots<i_{C_1}\leq A \\ 1\leq j_1<\cdots<j_{C_2}\leq C_2+C_3}} P_A(i_1,\cdots,i_{C_1})P_{C_2+C_3}(j_1,\cdots,j_{C_2}) \cr 
&\times\left[\left[\chi^{23}_{l_1}\otimes\chi^{123}_{l_2}\right]_L\otimes\left[N^{j_1\pi_1}(i_1,\cdots,i_{C_1})\otimes\left[N^{j_2\pi_2}(j_1,\cdots,j_{C_2})\otimes N^{j_3\pi_3}(j_{C_2+1},\cdots,j_{C_2+C_3})\right]_{j_{23}}\right]_{j_{123}}\right]_{JK}
\end{aligned}
\end{equation}
with the overlap kernels defined as
\begin{equation}
\begin{aligned}
    \chi^{23}_{l_1m_{l1}}(a_1;j_1,\cdots,j_{C_2+C_3})&=\left<\frac{\delta(r_1-a_1)}{r_1^2}Y_{l_1m_{l1}}(\hat{r}_1)\middle|\chi^{23}(\boldsymbol{r}_1;j_1,\cdots,j_{C_2+C_3})\right>  \cr
    \chi^{123}_{l_2m_{l2}}(a_2;i_1,\cdots,i_{C_1},j_1,\cdots,j_{C_2+C_3})&=\left<\frac{\delta(r_2-a_2)}{r_2^2}Y_{l_2m_{l2}}(\hat{r}_2)\middle|\chi^{123}(\boldsymbol{r}_2;i_1,\cdots,i_{C_1},j_1,\cdots,j_{C_2+C_3})\right>  \cr
    N^{j_1\pi_1}_{m_1}(i_1,\cdots,i_{C_1})&=\left<\Psi^{j_1\pi_1}_{m_1C_1}\middle|\Phi^\mathrm{int}_{C_1}(i_1,\cdots,i_{C_1})\right> \cr
    N^{j_2\pi_2}_{m_2}(j_1,\cdots,j_{C_2})&=\left<\Psi^{j_2\pi_2}_{m_2C_2}\middle|\Phi^\mathrm{int}_{C_2}(j_1,\cdots,j_{C_2})\right>  \cr
    N^{j_3\pi_3}_{m_3}(j_{C_2+1},\cdots,j_{C_2+C_3})&=\left<\Psi^{j_3\pi_3}_{m_3C_3}\middle|\Phi^\mathrm{int}_{C_3}(j_{C_2+1},\cdots,j_{C_2+C_3})\right>.
\end{aligned}
\end{equation}
\end{widetext}

\section{Applications}
\label{app}

\begin{figure}[!tb]
\includegraphics
  [width=0.9\hsize]
  {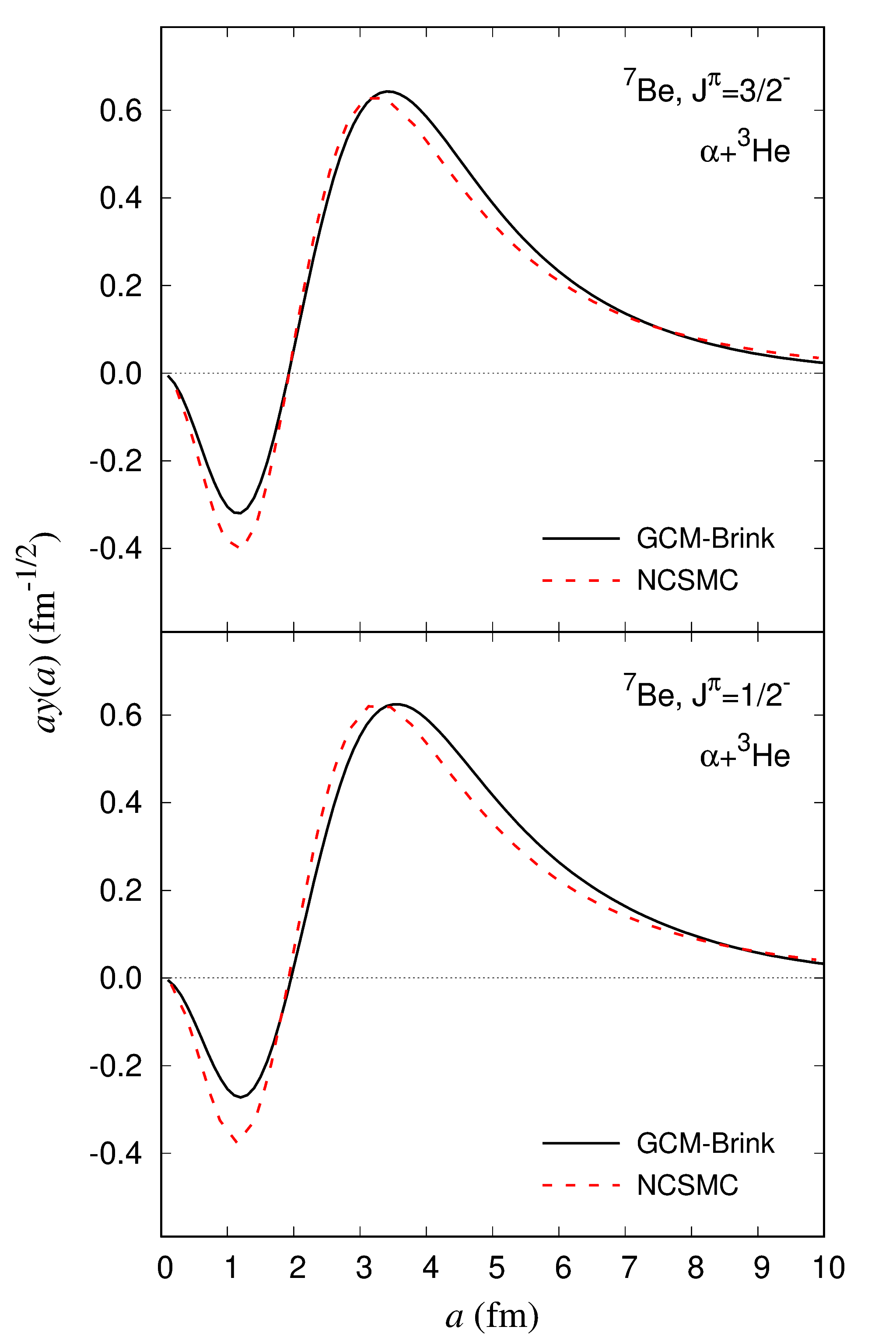}
\caption{The $\alpha+{}^3\mathrm{He}$ RWA for $3/2^-$ and $1/2^-$ states of $^7\mathrm{Be}$ calculated by GCM and NCSMC~\cite{vorabbi2019PRC024304}.}
\label{fig:rwa_Be7}
\end{figure}

The RWA, or cluster form factor in NCSM formalism, or overlap function in reaction theories, is extensively applied in a wide range of studies.
Beyond analyzing cluster configurations and calculating reaction cross sections, the RWA (cluster form factor) also serves as an indispensable quantity for solving the equation of motion in the no-core shell model with continuum (NCSMC), an \textit{ab initio} theory combining the NCSM with the cluster model~\cite{baroni2013PRL022505,baroni2013PRC034326}.
As shown in Fig.~\ref{fig:rwa_Be7}, the RWA calculated by GCM and NCSMC~\cite{vorabbi2019PRC024304} are compared for the $\alpha+{}^3\mathrm{He}$ structure in the ground and the first excited states of $^7\mathrm{Be}$.
We can see that, despite the quite different model wave function and interaction adopted, the results obtained by NCSMC are basically consistent with those obtained by GCM, especially for the tail part.
Besides, the positions of nodes that correspond to the forbidden states predicted by these two theoretical models also agree very well with each other.
The main distinction between the results from these two models is that, compared with those obtained from GCM, the NCSMC results exhibit higher inner peaks and more inward surface peaks.

In the present article, we focus on reviewing the theory and applications of RWA in nuclear clustering studies.
In the following part, we will demonstrate the applications of RWA in the analyses of cluster structures based on cluster-model wave functions.
Additionally, some reaction studies closely related to the cluster structures in nuclei and the calculation of RWA will also be briefly discussed.

\subsection{Clustering structure in $N\alpha$ nuclei}

One of the most interesting phenomena relating to clustering in the nuclei is the existence of the Hoyle state and its analogs in $N\alpha$ self-conjugate nuclei, in which the $\alpha$ clusters simultaneously occupy the lowest $0^+$ state and present a characteristic of Bose-Einstein condensation~(BEC)~\cite{tohsaki2001PRL192501}. 
The Hoyle state is essential for the evolution of life as it plays a crucial role in the nucleosynthesis of isotopes heavier than helium~\cite{hoyle1954AJSS121,cook1957PR508}. 
However, the structural configuration of the Hoyle state has been debated for decades since its discovery~\cite{morinaga1956PR254,morinaga1966PL78,suzuki1972PTP1517}.
It will be shown in the subsequent discussions that the RWA is an effective tool for searching and verifying the $N\alpha$ condensation states in light nuclei.

\begin{figure*}[!tb]
\includegraphics
  [width=\hsize]
  {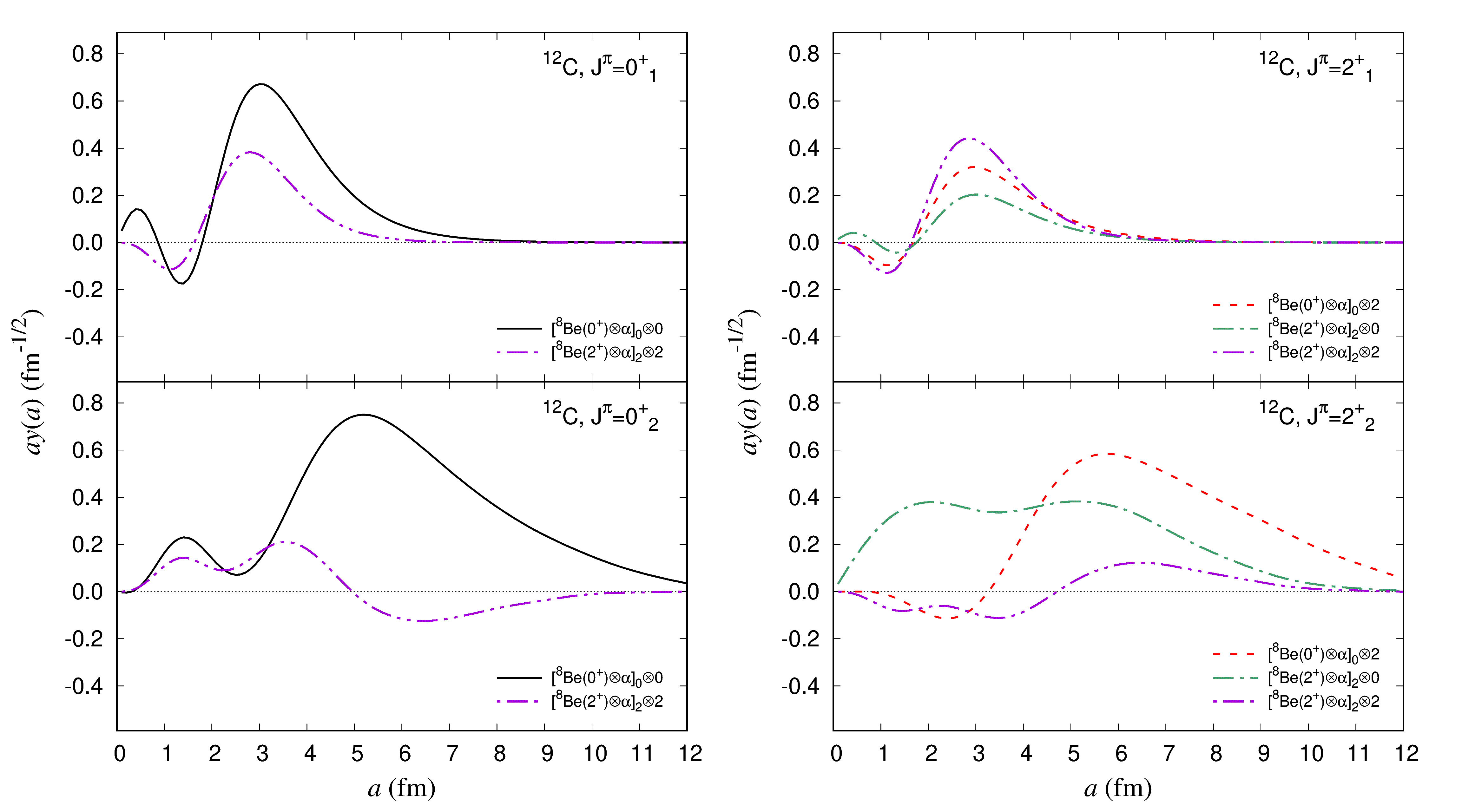}
\caption{The calculated RWA for various $^8\mathrm{Be}+\alpha$ channels in $0^+_1$, $0^+_2$, $2^+_1$, and $2^+_2$ states of $^{12}\mathrm{C}$.}
\label{fig:rwa_C12}
\end{figure*}

\begin{figure}[!tb]
\includegraphics
  [width=\hsize]
  {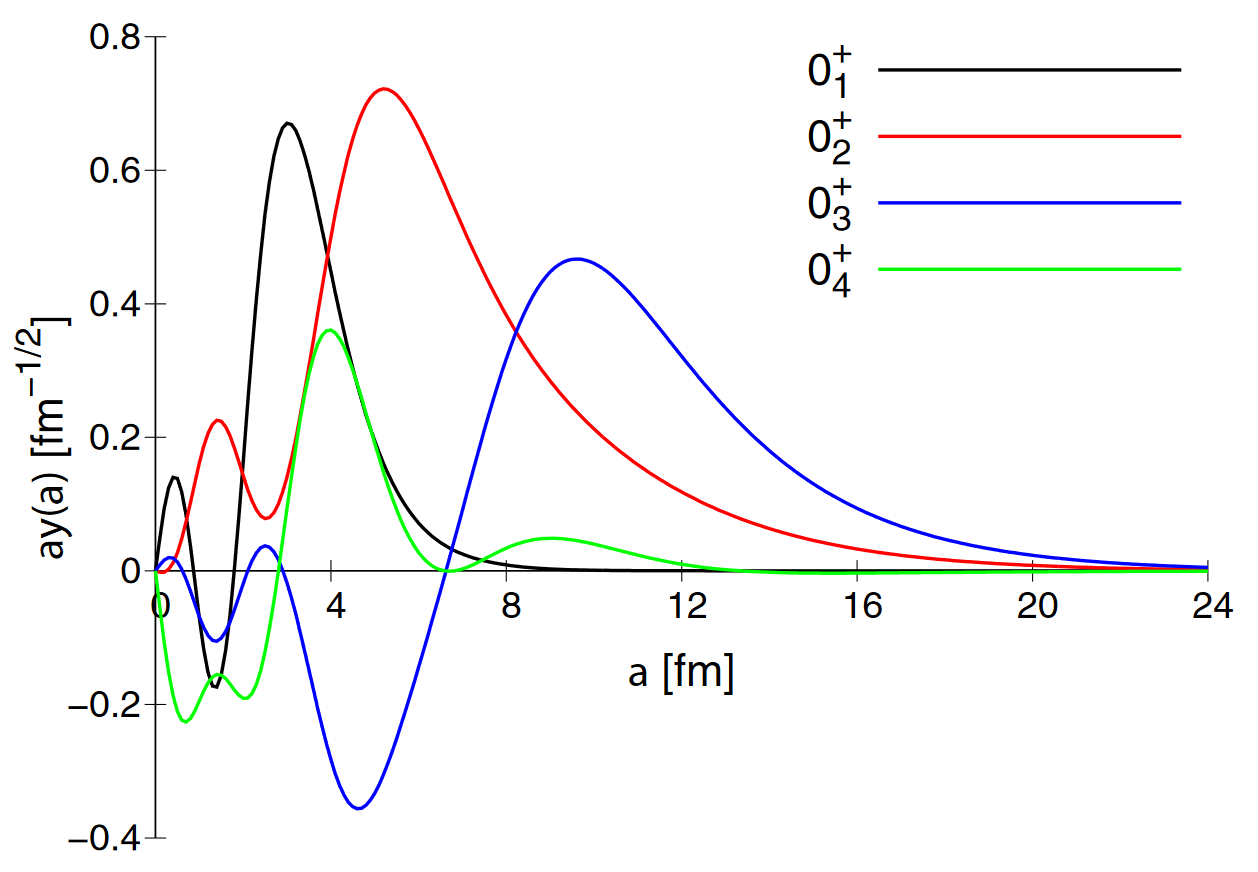}
\caption{The RWA of $[^8\mathrm{Be}(0^+)\otimes\alpha]_0\otimes0$ channel in $0^+_1$, $0^+_2$, $0^+_3$, and $0^+_4$ states of $^{12}\mathrm{C}$. The Figure is taken from Ref.~\cite{zhou2016PRC044319}.}
\label{fig:rwa_C12_0+}
\end{figure}

By means of $\alpha+\alpha+\alpha$ GCM, Uegaki et al.~\cite{uegaki1979PTP1621} conducted a systematic calculation of the ground and excited states of $^{12}\mathrm{C}$.
To study the coupling between the $\alpha$ cluster and the $^8\mathrm{Be}$ in the ground or excited states, the RWA of various $^8\mathrm{Be}+\alpha$ channels were evaluated for the obtained states of $^{12}\mathrm{C}$.
It was found that the Hoyle state, $0^+_2$, is predominantly composed of the channel $[^8\mathrm{Be}(0^+)\otimes\alpha]_0\otimes0$.
As $^8\mathrm{Be}$ is a well-known weakly-bound nucleus, Uegaki et al.\ concluded that the Hoyle state is constructed by the weak coupling of $^8\mathrm{Be}$ and $\alpha$ clusters or equivalently three $\alpha$ clusters, suggesting that the three $\alpha$ clusters are all occupying the lowest $0S$ orbit and forming a gaslike state. 
A few years later, such a result has been confirmed by various subsequent studies on the Hoyle state~\cite{kamimura1981NPA456,descouvemont1987PRC54}.
We present in Fig.~\ref{fig:rwa_C12} our results of the RWA for various $^8\mathrm{Be}+\alpha$ channels in $0^+_{1,2}$ and $2^+_{1,2}$ states of $^{12}\mathrm{C}$ obtained by GCM.
For the lowest two states, $0^+_1$ and $2^+_1$, the clustering structures contain more than one channel, with comparative amplitudes and similar curve structures.
This character, a mixture of multiple cluster configurations, together with the typical inner oscillations and the short tails, suggests a more pronounced shell-model structure of these two states.
In contrast, the $0^+_2$ and $2^+_2$ states exhibit more considerable clustering features.
As noted by Uegaki et al., compared with the ground state, the RWA of $[^8\mathrm{Be}(0^+)\otimes\alpha]_0\otimes0$ channel in $0_2^+$ state is significantly enhanced in amplitude.
The outward-shifted peak is also consistent with the larger radius of the Hoyle state.
The clustering structure of the $2^+_2$ state is more complicated, dominated by the $[^8\mathrm{Be}(2^+)\otimes\alpha]_0\otimes0$ and $[^8\mathrm{Be}(0^+)\otimes\alpha]_0\otimes2$ channels for the interior and exterior regions, respectively.
More recently, the $0^+$ states higher than the Hoyle state were reanalyzed using the THSR wave function to search for correlations between $\alpha$ clusters~\cite{zhou2016PRC044319}. 
By calculating the $[^8\mathrm{Be}(0^+)\otimes\alpha]_0\otimes0$ RWA, as shown in Fig.~\ref{fig:rwa_C12_0+}, the $0^+_3$ state was recognized as a breathing-like excited state of the Hoyle state, as it exhibits a very extended amplitude and one more node than the Hoyle state in the RWA, while the $0^+_4$ state is considered to be a possible bent-arm-structure state, with a much suppressed $[^8\mathrm{Be}(0^+)\otimes\alpha]_0\otimes0$ amplitude in the RWA. 

\begin{figure}[!tb]
\includegraphics
  [width=0.8\hsize]  
  {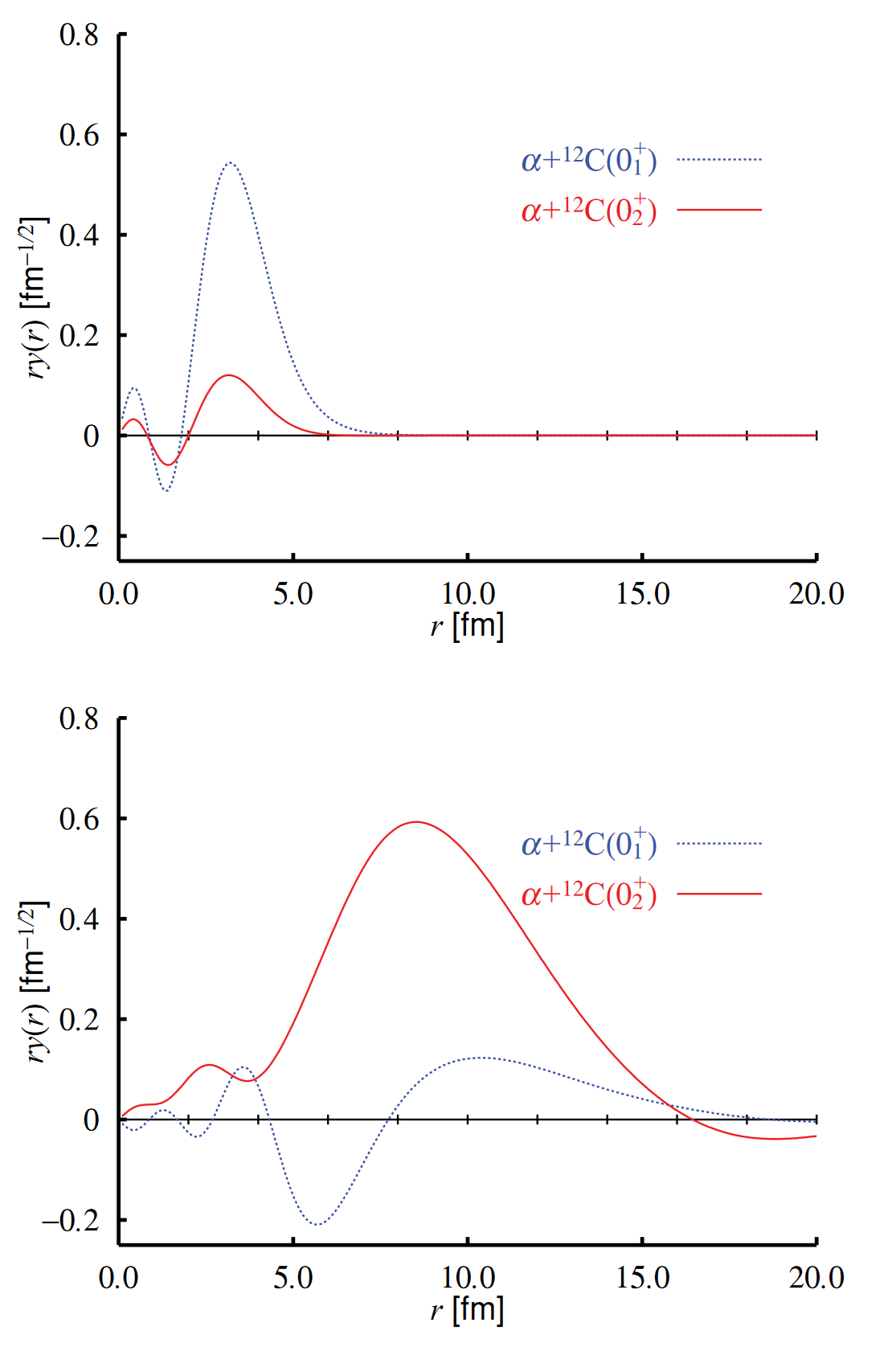}
\caption{The RWA of $[^{12}\mathrm{C}(0^+_{1,2})\otimes\alpha]_0\otimes0$ channel in $0^+_1$ and $0^+_4$ states of $^{16}\mathrm{O}$ obtained by THSR calculation. The Figure is taken from Ref.~\cite{funaki2010PRC024312}.}
\label{fig:rwa_O16}
\end{figure}

Although the $4\alpha$-condensate state of $^{16}\mathrm{O}$ has been predicted using the gaslike THSR wave functions~\cite{tohsaki2001PRL192501}, the identification of this state was controversial for years. 
With the semi-microscopic method OCM, Funaki et al.~\cite{funaki2008PRL082502} reproduced the full experimental spectrum of $0^+$ states for $^{16}\mathrm{O}$ up to $0^+_6$ (denoted as $(0^+_6)_\mathrm{OCM}$ hereafter), giving the RWA results of various channels for the $(0^+_6)_\mathrm{OCM}$ state, which was considered in that work as a strong candidate for the Hoyle-analog $4\alpha$-condensate state.
The calculated RWA clearly showed that the obtained $(0^+_6)_\mathrm{OCM}$ state is predominantly composed of the $[^{12}\mathrm{C}(0^+_2)\otimes\alpha]_0\otimes0$ channel with an extended tail, while the ground state is dominated by the channel with $^{12}\mathrm{C}(0_1^+)$ and has a steeply decreasing tail. 
Later, the same authors~\cite{funaki2010PRC024312} more elaborately analyzed the $0^+$ states of $^{16}\mathrm{O}$ employing the THSR wave functions.
Interestingly, in that work, the state with consistent characters as the $(0^+_6)_\mathrm{OCM}$ state, including the resonance energy, the r.m.s.\ radius, and $M(E0)$ transition strength, was identified as the fourth $0^+$ state, denoted as $(0^+_4)_\mathrm{THSR}$.
The correspondence of the structure between these two theoretically obtained states was further confirmed by comparing the RWA results.
The RWA of $\alpha+{}^{12}\mathrm{C}$ configurations calculated for $(0^+_4)_\mathrm{THSR}$ (Fig~\ref{fig:rwa_O16}) was found to exhibit similar features to that of $(0^+_6)_\mathrm{OCM}$.
The reason for the scarcity of the states obtained by the THSR method can be understood considering that complex channels involving higher-angular-momentum or negative-parity states of $^{12}\mathrm{C}$ were not included in that early version of the THSR wave function, which was designed to describe the gaslike structures.

\begin{figure}[!tb]
\includegraphics
  [width=\hsize]
  {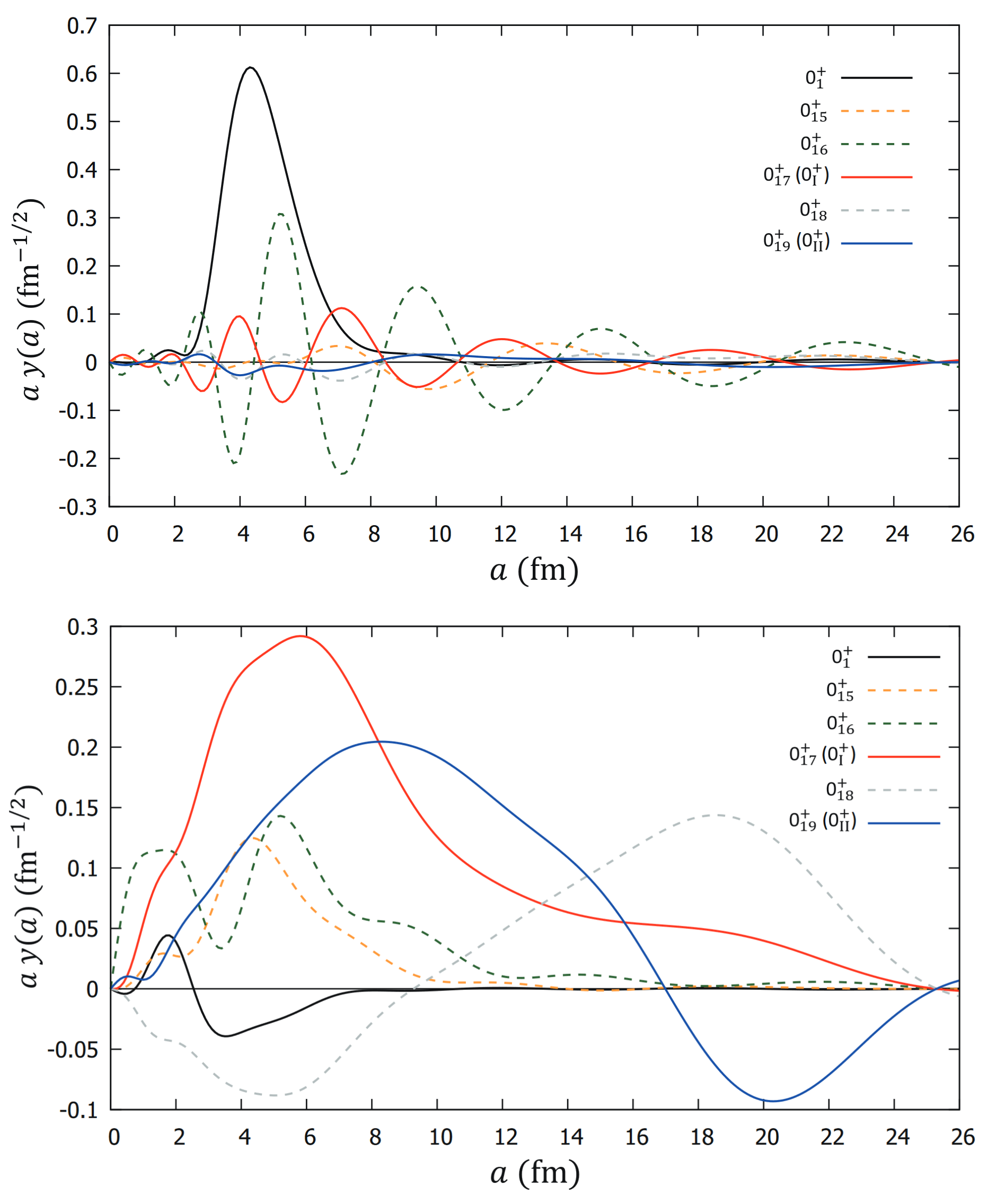}
\caption{The RWA of $[^{16}\mathrm{O}(0^+_1)\otimes\alpha]_0\otimes0$ and  $[^{16}\mathrm{O}(0^+_6)\otimes\alpha]_0\otimes0$ channels in ground and excited states above $5\alpha$ threshold of $^{20}\mathrm{Ne}$. The Figure is taken from Ref.~\cite{zhou2023NC8206}.} 
\label{fig:rwa_Ne20}
\end{figure}

Consisting of five $\alpha$ clusters, the relative motions between clusters in $^{20}\mathrm{Ne}$ are more complicated.
By calculating the $^{16}\mathrm{O}+\alpha$ RWA for various rotational bands of $^{20}\mathrm{Ne}$, Kimura~\cite{kimura2004PRC044319} investigated the relative motions between the $^{16}\mathrm{O}$ and $\alpha$ clusters.
Similar to $^{12}\mathrm{C}$ and $^{16}\mathrm{O}$, for low-lying states of $^{20}\mathrm{Ne}$, the RWA results of various $^{16}\mathrm{O}+\alpha$ components are oscillating and suppressed in the interior region, and are obviously enhanced in the exterior region.
From the calculated RWA, it is also demonstrated that the ground band of $^{20}\mathrm{Ne}$ exhibits a pronounced anti-stretching phenomenon, since the average distance between the $^{16}\mathrm{O}$ and $\alpha$ clusters decreases as the angular momentum increases.
The $5\alpha$ condensate state is predicted to lie much higher, about $20~\mathrm{MeV}$, making the search for this state much more difficult.
Quite recently, Zhou et al.~\cite{zhou2023NC8206} theoretically recognized the $5\alpha$ condensate state of $^{20}\mathrm{Ne}$ at $2.7~\mathrm{MeV}$ above the $5\alpha$ threshold.
The calculated RWA shows that, analogous to the $3\alpha$ and $4\alpha$ condensate states, the $5\alpha$ condensate state (denoted as $0^+_\mathrm{I}$) is dominated by the $[^{16}\mathrm{O}(0^+_6)\otimes\alpha]_0\otimes0$ channel, with features of an enhanced and extended amplitude and zero node, as depicted in Fig.~\ref{fig:rwa_Ne20}.
Additionally, another higher $0^+$ state, denoted as $0^+_\mathrm{II}$, was also found to possess a significant $[^{16}\mathrm{O}(0^+_6)\otimes\alpha]_0\otimes0$ component, but with one node in the RWA.
The enhanced monopole transition between $0^+_\mathrm{I}$ and $0^+_\mathrm{II}$, as well as the nodal structures of their RWA, suggests that the $0^+_\mathrm{II}$ state is the breathing-like excitation of the $5\alpha$ condensate state.

\subsection{Clustering in neutron-rich nuclei}

Apart from the Hoyle and Hoyle-analog states in $N\alpha$ nuclei, the neutron-rich nuclei also exhibit various interesting cluster states~\cite{liu2018NST184}.
Since the 1990s, both experimental~\cite{korsheninnikov1995PLB53} and theoretical studies~\cite{vonoertzen1996ZPA37,vonoertzen1997ZPA355,arai2004PRC014309} have suggested that a novel form of clustering, the molecular structure, may exist in neutron-rich Be or C isotopes, in which the $\alpha$ clusters interplay with the extra neutrons, serving as valence neutrons akin to covalent electrons binding the atoms in a molecule.
One of the simplest examples of a molecular state is $^9\mathrm{Be}$, where the unbound system of $\alpha+\alpha$ is bound by adding one valence neutron.
Combining the resonant state method with RGM, Arai et al.~\cite{arai2003PRC014310} systematically studied the structure of the ground and excited states of $^9\mathrm{Be}$, calculating the $^8\mathrm{Be}+n$ and $^5\mathrm{He}+\alpha$ RWA for some low-lying states, including $3/2^-$, $5/2^-$, $1/2^+$, $1/2^-$, and $5/2^+$.
The results show that in these states, both the $^8\mathrm{Be}+n$ and $^5\mathrm{He}+\alpha$ configurations have large amplitudes.
Besides, the maximum amplitude is reached with a larger distance between clusters for positive-parity states compared to negative-parity ones, suggesting that the average $\alpha$-$\alpha$ distances in positive-parity states are larger, consistent with earlier studies~\cite{okabe1979PTP1049}.

\begin{figure}[!tb]
\includegraphics
  [width=\hsize]
  {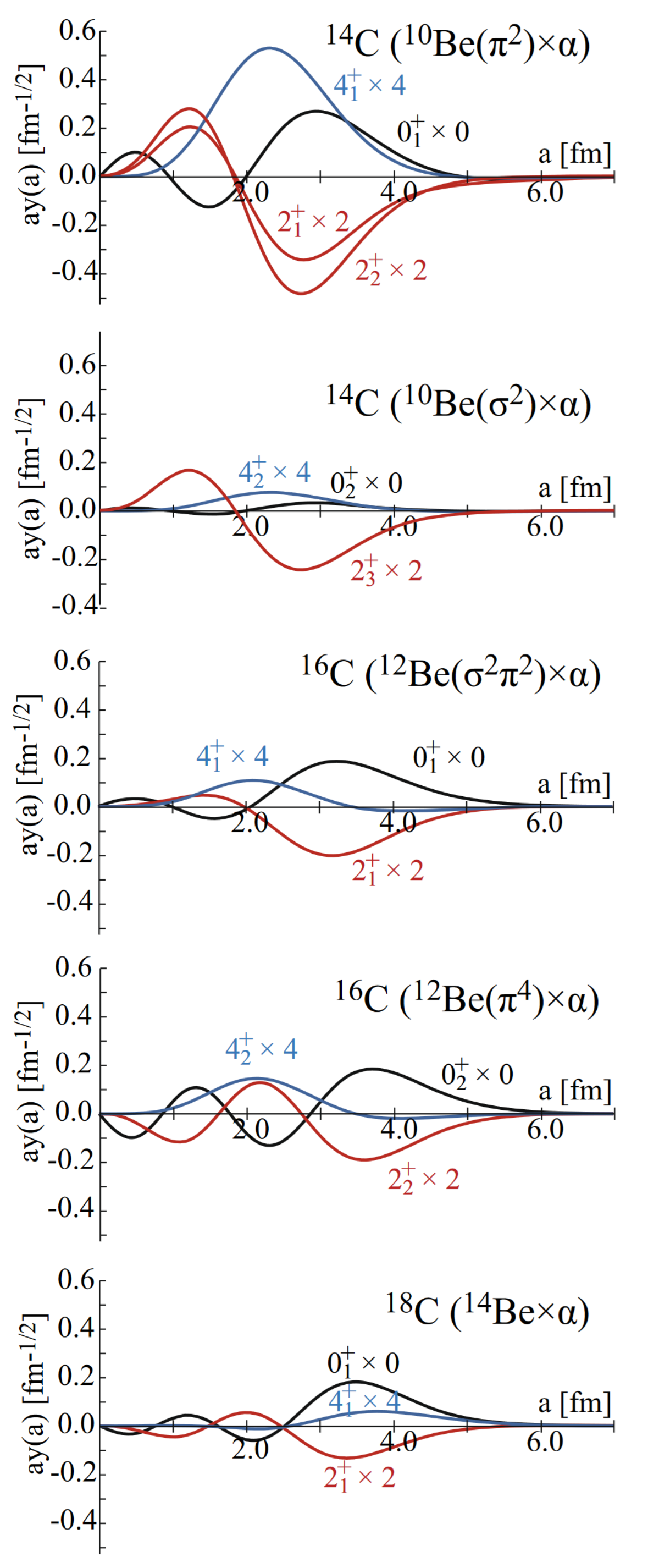}
\caption{The RWA of $^{X-4}\mathrm{Be}+\alpha$ channels in ground states of $^{14}\mathrm{C}$, $^{16}\mathrm{C}$, and $^{18}\mathrm{C}$. The Figure is taken from Ref.~\cite{zhao2021TEPJA157}.}
\label{fig:rwa_C14-18}
\end{figure}

More structural information, other than binary clustering probability, can be inferred from the RWA of neutron-rich nuclei.
For example, the orbits of valence neutrons were analyzed in a pioneering study~\cite{kanada-enyo2003PRC014319} on molecular states in $^{12}\mathrm{Be}$.
By analyzing the the $^6\mathrm{He}+{}^6\mathrm{He}$ and  $^8\mathrm{He}+\alpha$ RWA of different rotational bands, and combining this with the picture of molecular orbits of neutrons, Kanada-En'yo and Horiuchi concluded that in the $K^\pi=0^+_1$ band, the neutrons are surrounding the core consisting of the two $\alpha$ clusters, as both the $^6\mathrm{He}+{}^6\mathrm{He}$ and  $^8\mathrm{He}+\alpha$ RWA in these states are significant.
On the other hand, the neutrons move around either one of the $\alpha$ clusters in the $K^\pi=0^+_3$ band, provided that only the $^6\mathrm{He}+{}^6\mathrm{He}$ structure dominates these states.
Besides, the tendency of $\alpha$ clustering as the drip-line approaches is another interesting problem that can be studied by analyzing RWA. 
A recent experiment demonstrated a negative correlation between $\alpha$ formation and the thickness of the neutron-skin for Sn isotopes~\cite{tanaka2021S260}.
Later, the $\alpha$ cluster formation in neutron-rich isotopes of Be and B was systematically analyzed by calculating the RWA of $\alpha+{}^{X-4}\mathrm{He}$ structures~\cite{motoki2022PoTaEP113D01}.
It was found that, although the $\alpha$ RWA reduced as the neutron number increased, the total cluster formation probability is still enhanced, since the structures consisting of neutron-rich clusters, such as $^6\mathrm{He}+{}^6\mathrm{He}$ in $^{12}\mathrm{Be}$ and $^6\mathrm{He}+{}^8\mathrm{He}$ in $^{14}\mathrm{Be}$, become comparable with the $\alpha$-clustering structures. 
The analogous relationship between the $\alpha$ formation in the nuclear surface and the richness of the neutrons has also been studied for C isotopes~\cite{zhao2021TEPJA157}.
In that work, the RWA of various $\alpha+{}^{X-4}\mathrm{Be}$ channels were calculated for the ground states of $^{14}\mathrm{C}$, $^{16}\mathrm{C}$, and $^{18}\mathrm{C}$, as shown in Fig~\ref{fig:rwa_C14-18}.
The results show that for $^{14}\mathrm{C}$, the $\alpha$ amplitude in the exterior is large and comparable with that for $^{12}\mathrm{C}$, while for $^{16}\mathrm{C}$ and $^{18}\mathrm{C}$, with thicker neutron-skins, the $\alpha$ formation in the nuclear surface is considerably suppressed.

\begin{figure}[!tb]
\includegraphics
  [width=0.9\hsize]
  {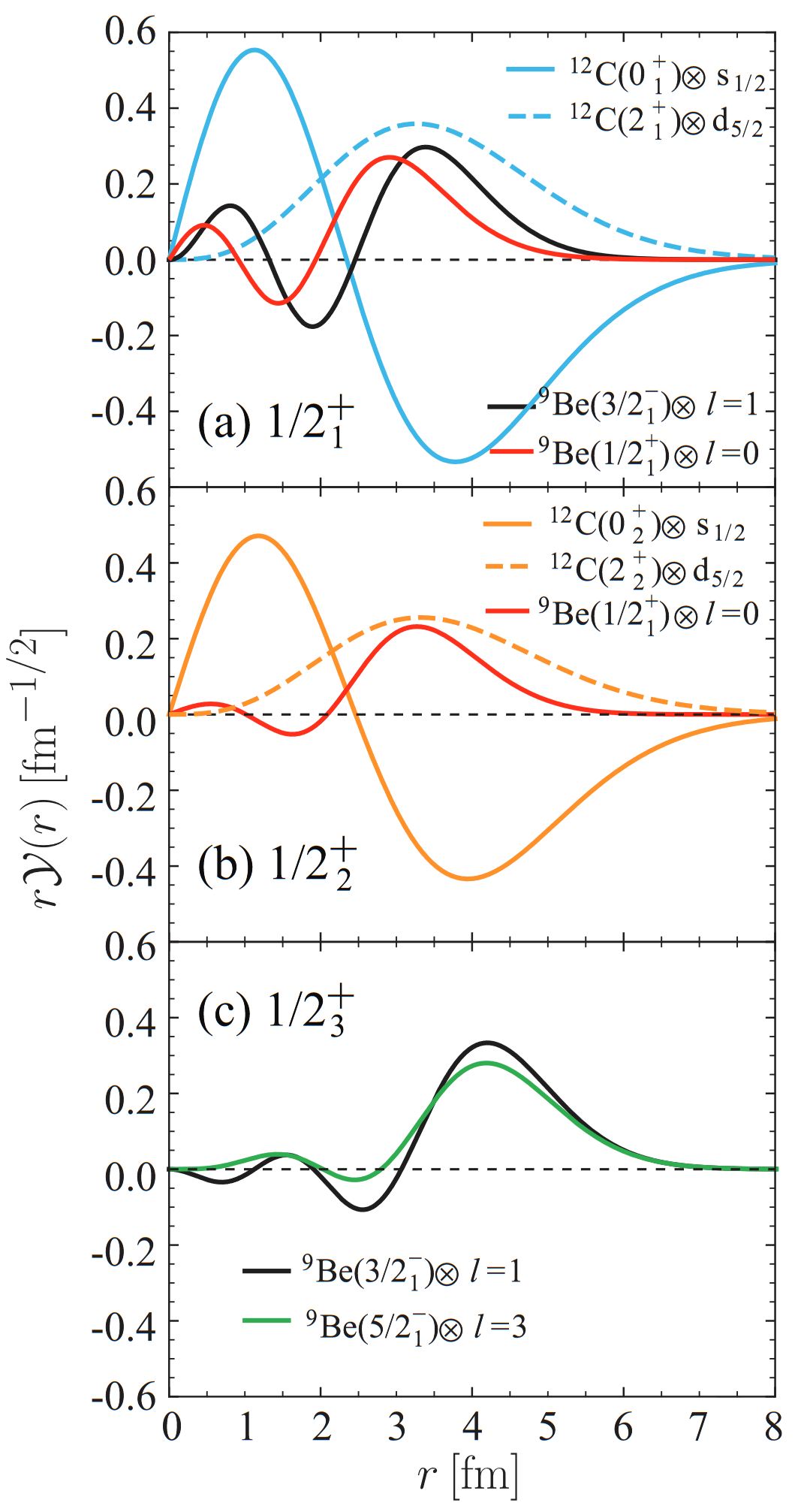}
\caption{The RWA of various $^{12}\mathrm{C}+n$ and $^{9}\mathrm{Be}+\alpha$ channels in $1/2^+_1$, $1/2^+_2$, and $1/2^+_3$ states of $^{13}\mathrm{C}$. The Figure is taken from Ref.~\cite{chiba2020PRC024317}.}
\label{fig:rwa_C13}
\end{figure}

\begin{figure}[!tb]
\includegraphics
  [width=\hsize]
  {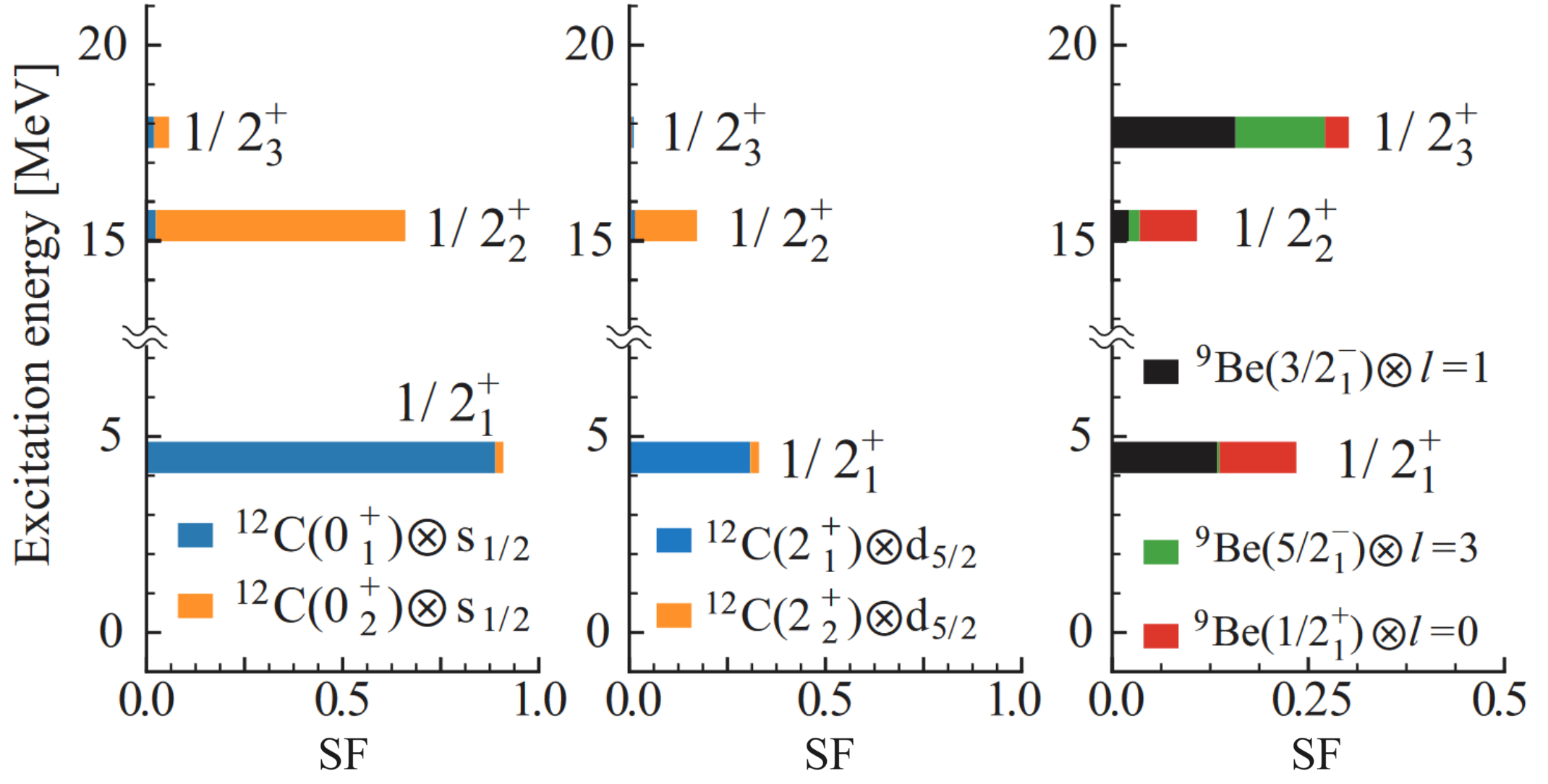}
\caption{The SF of $^{12}\mathrm{C}+n$ and $^{9}\mathrm{Be}+\alpha$ channels in $1/2^+_1$, $1/2^+_2$, and $1/2^+_3$ states of $^{13}\mathrm{C}$. The Figure is taken from Ref.~\cite{chiba2020PRC024317}.}
\label{fig:sf_C13}
\end{figure}

In neutron-rich nuclei, the formation of exotic clusters other than $\alpha$ is drawing increasing recent interest.
By the use of RWA, the formation probabilities of not only the $\alpha$ cluster but also light exotic clusters such as triton, $^3\mathrm{He}$, deuteron, etc., can be theoretically estimated. 
More significantly, another kind of Hoyle-analog state, composed of different kinds of clusters forming a gaslike state, has been proposed to exist among these exotic clustering states. 
A natural candidate for searching for non-$N\alpha$ Hoyle-analog states is $^{11}\mathrm{B}$, which can be well described by the $\alpha+\alpha+t$ cluster model.
By means of AMD, Kanada-En'yo~\cite{kanada-enyo2007PRC024302} found that the $3/2^-_3$ state of $^{11}\mathrm{B}$ is charactered by a dilute density distribution, similar to that of the Hoyle state.
Years later, Yamada and Funaki~\cite{yamada2010PRC064315} pointed out that the essence of the Hoyle-analog state lies in the occupancy of the lowest $0S$ orbit for all the constituent clusters, and thus to avoid the Pauli-blocking effect among them to form a gaslike configuration.
By OCM calculation, they demonstrated that the $1/2^+_2$ state of $^{11}\mathrm{B}$ should be a strong candidate for the Hoyle-analog state of $\alpha+\alpha+t$ clustering.
Thus, due to the lack of experimental results, the identification of the Hoyle-analog state in $^{11}\mathrm{B}$ remains controversial.
Recent GCM calculation results of $^{11}\mathrm{B}$~\cite{zhou2018PRC054323} support the gaslike nature of the $3/2^-_3$ state, but are quite different in the description of the $1/2^+_2$ state, finding it exhibits a linear-chain-like structure.
Adding an extra nucleon to the Hoyle state is another direction for exploring the non-$N\alpha$ Hoyle-analog states.
The configuration of $^{12}\mathrm{C}(0^+_2)+n$ for $^{13}\mathrm{C}$ has been analyzed by OCM~\cite{yamada2015PRC034326} and AMD~\cite{chiba2020PRC024317}, and their predictions of the Hoyle-analog state are consistently $1/2^+_2$. 
The RWA obtained by AMD calculations are presented in Fig.~\ref{fig:rwa_C13}, in which we can see that the $[^{12}\mathrm{C}(0^+_2)\otimes n]_{1/2}\otimes0$ channel is predominant in the $1/2^+_2$ state, while the other $1/2^+$ states exhibit multi-channel configurations.
Furthermore, the SF of various $^{12}\mathrm{C}+n$ and $^9\mathrm{Be}+\alpha$ channels are also displayed in Fig.~\ref{fig:sf_C13}, and as expected, an obvious enhancement of the $[^{12}\mathrm{C}(0^+_2)\otimes n]_{1/2}\otimes0$ SF can be observed.

\begin{figure*}[!tb]
\includegraphics
  [width=\hsize]
  {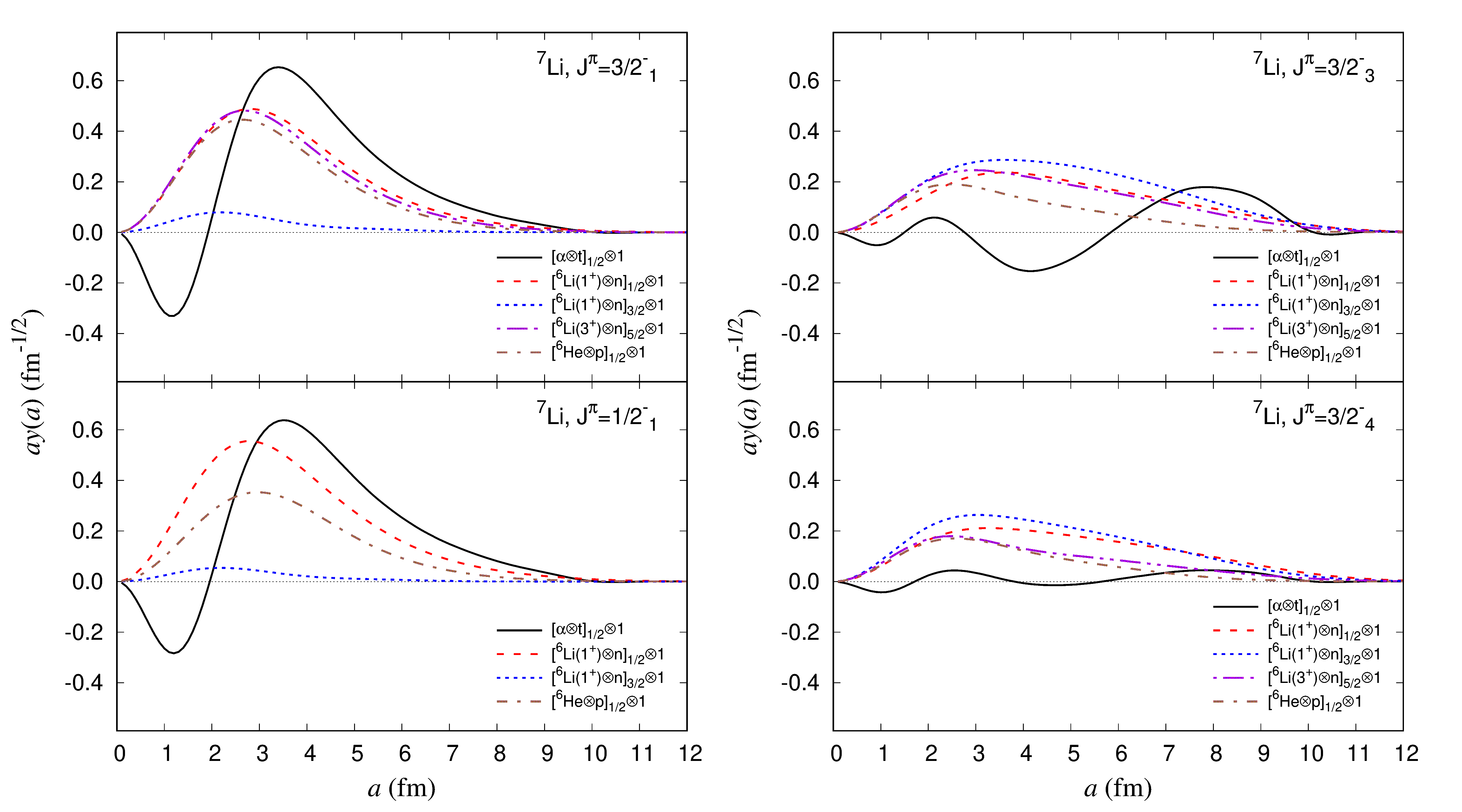}
\caption{The calculated RWA of $\alpha+t$, $^6\mathrm{Li}+n$, and $^6\mathrm{He}+p$ channels in $3/2^-_1$, $1/2^-_1$, $3/2^-_3$, and $3/2^-_4$ states of $^{7}\mathrm{Li}$.}
\label{fig:rwa_Li7}
\end{figure*}

For some high-lying states of light nuclei, the $\alpha$ cluster may interact with loosely bound valence nucleons to form various exotic clustering structures.
Taking the $^7\mathrm{Li}$ as an example, though the most well-known cluster structure of $\alpha+t$ in $^7\mathrm{Li}$ has been extensively studied for decades~\cite{mihailovic1978NPA377,toshitaka1984NPA185,kaneko1986PRC771}, the $\mathrm{core}+N$ configurations, including $^6\mathrm{Li}+n$ and $^6\mathrm{He}+p$, also constitute significant components in the ground state and even dominate in some highly excited states.
More importantly, these higher-lying states related to the $^6\mathrm{He}$ or $^6\mathrm{Li}$ channels may play roles in the production and destruction of $^7\mathrm{Li}$ in the early universe, which is tightly associated with the unsolved cosmological lithium problem~\cite{meissner2022EPJA212}.
Here, we calculated the RWA of various channels for some states of $^7\mathrm{Li}$ to show the variety of clustering configurations as the excitation energy increases.
The results are presented in Fig.~\ref{fig:rwa_Li7}.
From the results, we can clearly see the multi-channel character of the cluster configurations in the two bound states, the ground state $3/2^-$ and the first excited state $1/2^-$.
In these two states, apart from the dominant $\alpha+t$ configuration, the $^6\mathrm{He}+p$ and various $^6\mathrm{Li}+n$ channels also have significant amplitudes. 
This character may correspond to the mixture of the pronounced $\alpha+t$-cluster and the moderate shell-model configurations. 
On the other hand, the resultant RWA shows that in the two high-lying resonant states above the $\alpha+n+n+p$ threshold, $3/2^-_3$ and $3/2^-_4$, the $\alpha+t$ amplitude is drastically suppressed, especially for the higher $3/2^-_4$ state, while the $\mathrm{core}+N$ configurations, both the $^6\mathrm{He}+p$ and $^6\mathrm{Li}+n$ structures, become the dominant binary clustering structures.

\subsection{Analysis of the three-body correlations}

Recently, the correlations between clusters or nucleons are drawing increasing interests~\cite{huang2020PRC034615,ma2020PRC014910} and require analyses on the three-body motions of clusters.
The motion of two valence neutrons around the core relates closely to the well-known dineutron correlation~\cite{hagino2010JPG015105}, which is fundamental for more deeply understanding the nuclear force. 
Following the framework present in Sec.~\ref{subsec:3body_theor}, as examples, the two-body overlap amplitudes of the ground states of $^{6}\mathrm{He}$ and $^{10}\mathrm{Be}$ are calculated, depicting the relative motions for $\alpha+n+n$ and $^{8}\mathrm{Be}+n+n$ cluster structures, respectively.
From the results, the motions and correlations of two valence neutrons outside the core are clearly demonstrated, with $r_1$ denoting the distance between two valence neutrons, and $r_2$ denoting the distance from the core to the c.o.m.\ of the two valence neutrons.

\begin{figure}[!tb]
\includegraphics
  [width=\hsize]
  {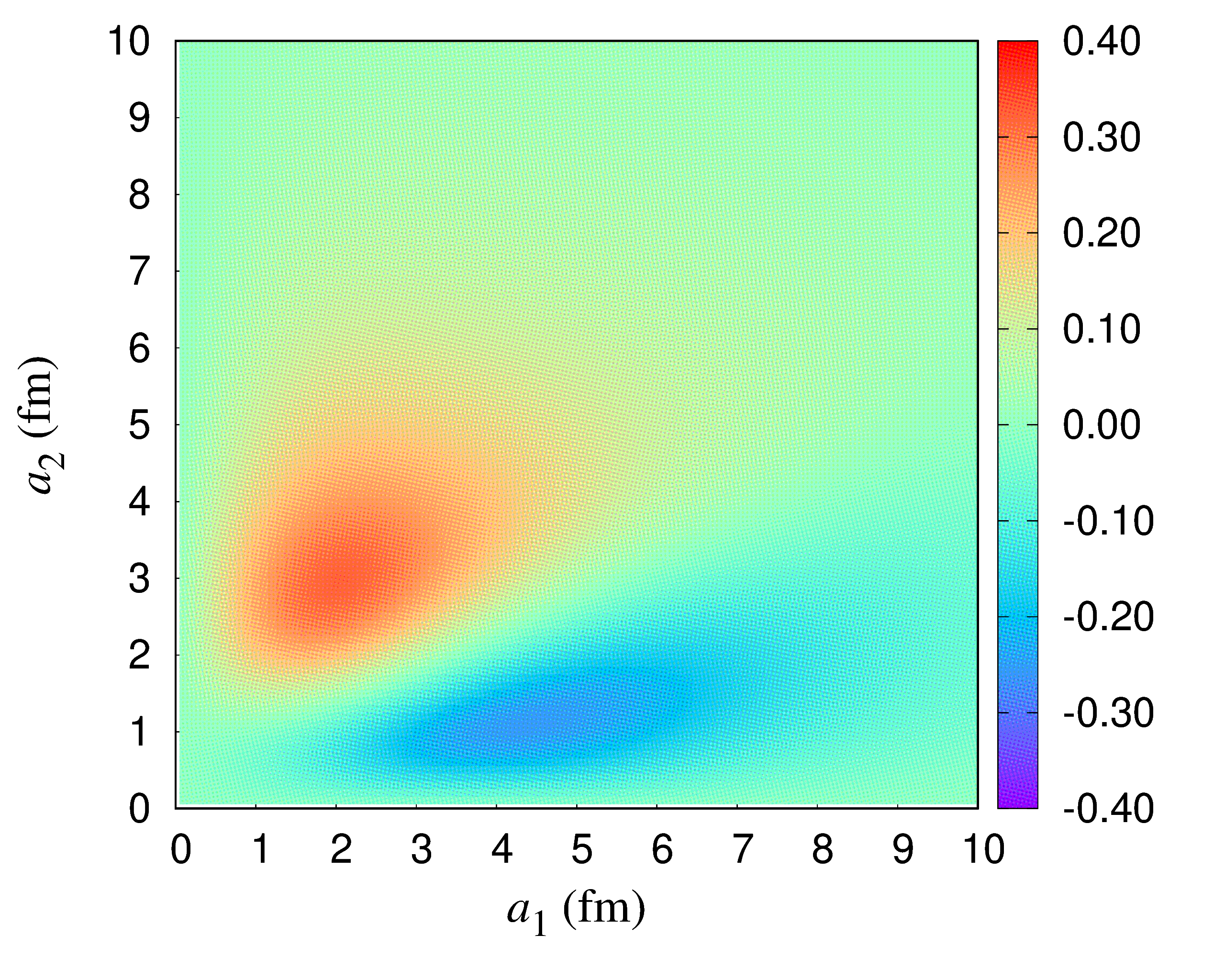}
\caption{The calculated two-body overlap amplitude of $[\alpha\otimes[n\otimes n]_0]_0\otimes[0\otimes0]_0$ channel in the ground state of $^{6}\mathrm{He}$.}
\label{fig:of3_He6}
\end{figure}

In Fig.~\ref{fig:of3_He6}, we show the two-body overlap amplitude of $[\alpha\otimes[n\otimes n]_0]_0\otimes[0\otimes0]_0$ configuration in $^6\mathrm{He}$.
The result exhibits two well-developed peaks, locating at $(r_1=4.4~\mathrm{fm}, r_2=1.1~\mathrm{fm})$ and $(r_1=2.1~\mathrm{fm},r_2=3.0~\mathrm{fm})$, respectively.
The peak in $(r_1=2.1~\mathrm{fm},r_2=3.0~\mathrm{fm})$ represents a stronger correlation between two valence neutrons than between valence neutron and the $\alpha$ core, which is always recognized as the dineutron configuration.
On the other hand, the smaller peak in $(r_1=4.4~\mathrm{fm}, r_2=1.1~\mathrm{fm})$ is associated with a ``cigar'' configuration, where the two valence neutrons are separated by larger distance, while their c.o.m.\ is close to the $\alpha$ core.
Lying between the two peaks is a narrow area where the overlap amplitude is close to $0$.
This area corresponds to the three-body forbidden states, similar to the zero-value nodes appearing in RWA curves.

\begin{figure}[!tb]
\includegraphics
  [width=\hsize]
  {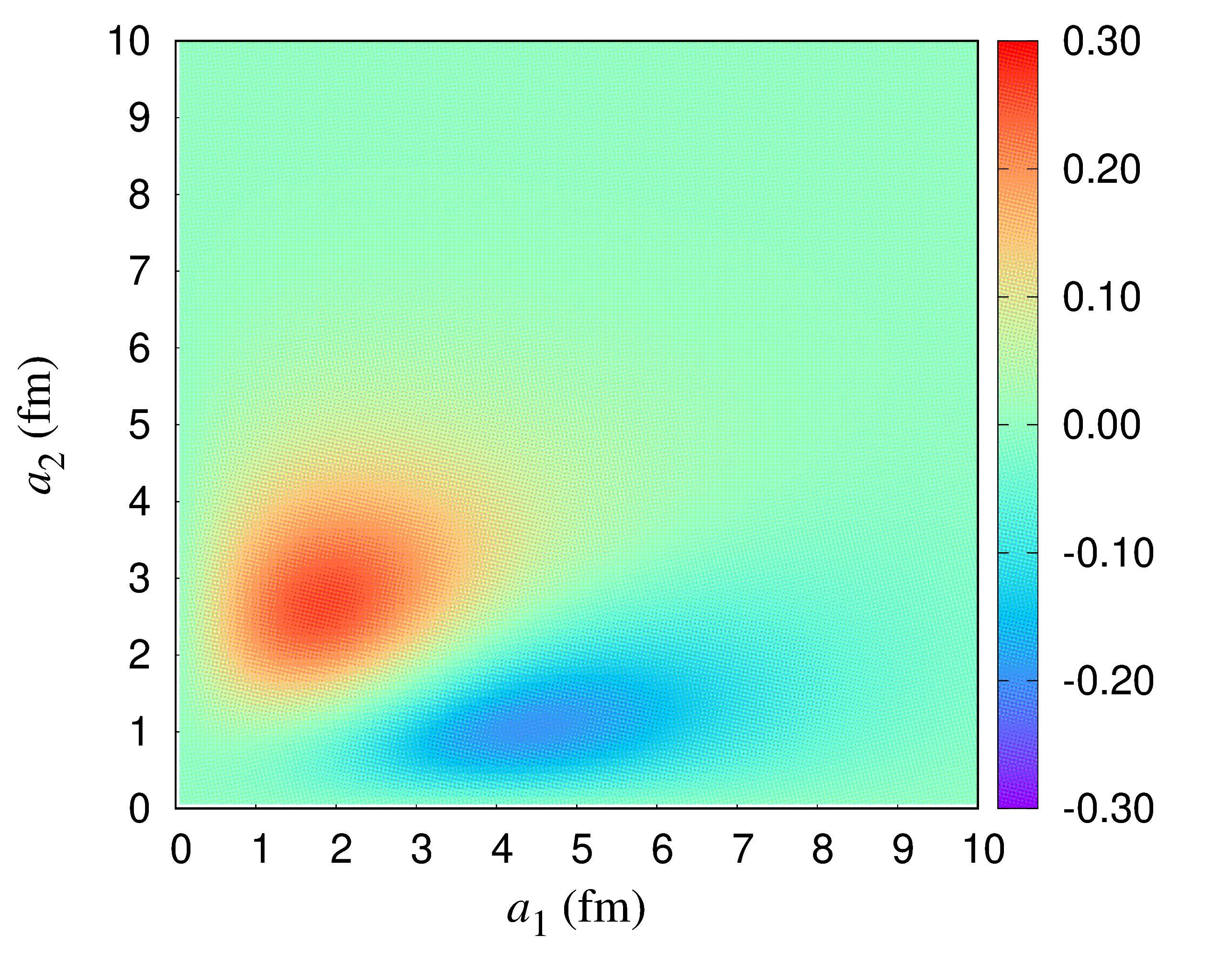}
\caption{The calculated two-body overlap amplitude of $[^8\mathrm{Be}\otimes[n\otimes n]_0]_0\otimes[0\otimes0]_0$ channel in the ground state of $^{10}\mathrm{Be}$.}
\label{fig:of3_Be10}
\end{figure}

To see the behavior of valence neutrons in the presence of a more complex core, we display the overlap amplitude of $^8\mathrm{Be}+n+n$ configuration in $^{10}\mathrm{Be}$ in Fig.~\ref{fig:of3_Be10}.
The same configuration has been analyzed before with a smaller basis space~\cite{kobayashi2016PRC024310}.
Here, we construct the basis space with more spatial configurations by taking into account the correlations between clusters as comprehensively as possible.
As the analogous structure of $\alpha+n+n$ in $^6\mathrm{He}$, the $^8\mathrm{Be}+n+n$ configuration also exhibits clearly separated areas associated with dineutron and cigar correlations, with the amplitude peaks at $(r_1=1.8~\mathrm{fm},r_2=2.7~\mathrm{fm})$ and $(r_1=4.4~\mathrm{fm},r_2=1.0~\mathrm{fm})$, respectively.
Compared to the $\alpha+n+n$ configuration, the amplitudes of $^8\mathrm{Be}+n+n$ are characterized by a more compact distribution, consistent with the smaller radius of $^{10}\mathrm{Be}$ than that of $^6\mathrm{He}$.
Notably, our results are in good agreement with the previous analysis of $^8\mathrm{Be}+n+n$ structure~\cite{kobayashi2016PRC024310}.

\subsection{Reaction cross section evaluation}

As a consequence of the development of microscopic structure studies, the microscopic structure information, including the RWA, is frequently adopted as an important input in recent reaction calculations.
The $\alpha$ knockout reaction is tightly related to the $\alpha$ clustering in nuclei, and thus, is extensively studied to probe cluster structures or to extract the experimental SF values~\cite{li2023PRL212501,mabiala2009PRC054612,yoshimura1998NPA3}.
Recently, the $^{20}\mathrm{Ne}(p,p\alpha)^{16}\mathrm{O}$ reaction was analyzed by Yoshida et al.~\cite{yoshida2019PRC044601}, where the $^{16}\mathrm{O}$-$\alpha$ cluster wave function was obtained from the RWA microscopically calculated.
Impressively good consistency between reaction and structure analyses was demonstrated, as the experimental data were reproduced well without including any additional correction or scaling during the calculation of the cross sections.

\begin{figure*}[!tb]
\includegraphics
  [width=0.9\hsize]
  {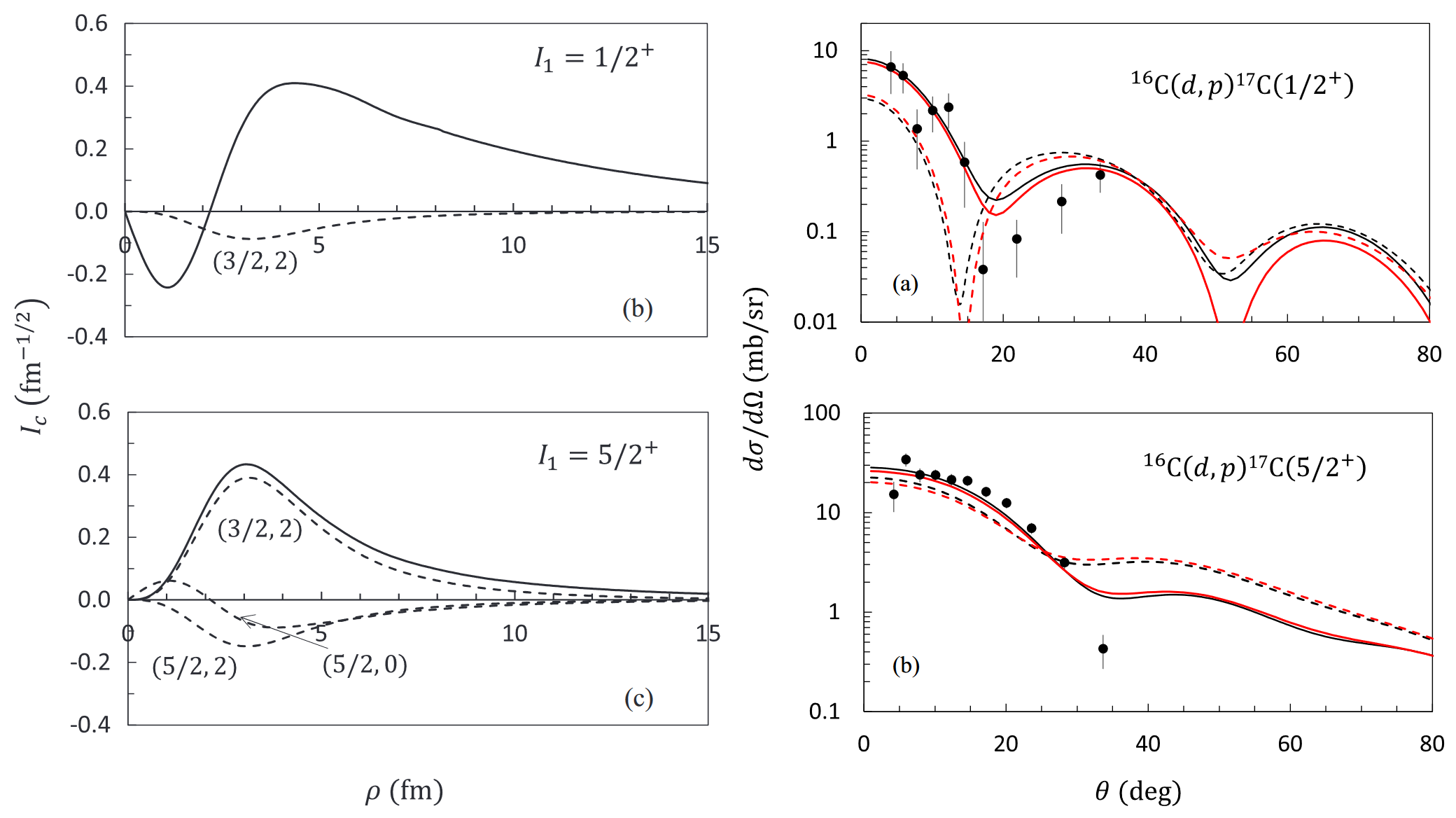}
\caption{The RWA of $^{16}\mathrm{C}+n$ channels in $1/2^+$ and $5/2^+$ states of $^{17}\mathrm{C}$, and the $^{16}\mathrm{C}(d,p){}^{17}\mathrm{C}$ transfer cross sections calculated using the RWA as input. The Figure is taken from Ref.~\cite{chien2023PRC044605}.}
\label{fig:cs_C17}
\end{figure*}

The transfer reactions is also usually applied to extract the structure information of nuclei~\cite{liu2020NST20}. 
By combining the CDCC and the DWBA improved by microscopic input of RWA, Chien and Descouvemont~\cite{chien2023PRC044605} studied the $^{16}\mathrm{C}(d,p)^{17}\mathrm{C}$ transfer reaction and calculated the cross section, as shown in Fig.~\ref{fig:cs_C17} along with the microscopically calculated RWA.
In this method, the wave functions of $^{16}\mathrm{C}$ and $^{17}\mathrm{C}$ are obtained by the microscopic cluster model, RGM, and the $^{16}\mathrm{C}+d$ and $^{17}\mathrm{C}+p$ scattering wave functions are calculated using optical potential, in which for the entrance channel, $^{16}\mathrm{C}+d$, the breakup effects of deuteron are simulated by the CDCC method.
Then, the cross sections are obtained by evaluating the transfer scattering matrix.
The calculated transfer cross sections, in comparison with the experimental data, demonstrate a good agreement between theoretical and experimental results.
Noticeably, during the calculation, no adjustable parameter is induced, and the results are insensitive to the choice of optical potential.

\section{Summary and Outlook}
\label{sum}

In this review, we briefly present the forms of various microscopic cluster-model wave functions while recalling some of the typical cluster models, including RGM, GCM, and OCM.
The significance of the reduced-width amplitude, also known as the cluster form factor or overlap function, is demonstrated from the point of view of structure and reaction analyses.
Based on the cluster-model wave functions, we present the definition and calculation methods of RWA.
Provided the GCM or THSR wave function constructed from the Brink cluster wave functions, the Laplace expansion method is an effective approach to calculate the RWA, compared to the traditional method.
Furthermore, the tail part of the RWA can be well approximated by calculating the overlap between the nuclear wave function and the single-Brink wave function.
Following the brief theoretical framework we revised, some application examples are presented to demonstrate the roles played by RWA in the structure and reaction analyses of light nuclei.
Additionally, we show the extension of the overlap-amplitude calculation from two-body structure analysis to three-body analysis for $^6\mathrm{He}$ ($\alpha+n+n$) and $^{10}\mathrm{Be}$ ($^8\mathrm{Be}+n+n$). 
By calculating the two-body overlap amplitude, some related studies for three-body correlations in nuclear structures are in progress. 




%

\end{document}